\documentclass[prb,superscriptaddress,nofootinbib,longbibliography,tightenlines]{revtex4-2}
\usepackage[svgnames]{xcolor}
\usepackage{amsmath,amssymb,amsthm,bbm}
\usepackage{wrapfig}
\usepackage{hyperref}
\usepackage{graphicx}
\usepackage{float}
\usepackage{braket}
\usepackage{dsfont}
\usepackage{import}
\usepackage{subcaption}  
\usepackage{amssymb}     
\usepackage{pifont} 
\usepackage[capitalise]{cleveref}
\hypersetup{
  colorlinks,
  citecolor=Salmon,
  linkcolor=Teal,
  urlcolor=Teal}
\usepackage{comment}  
\usepackage[normalem]{ulem}
\usepackage{soul}
\usepackage{svg}
\usepackage{orcidlink}

\usepackage{slashed}
\usepackage{subcaption}
\usepackage{ragged2e}
\DeclareCaptionJustification{justified}{\justifying}
\renewcommand{\i}{\mathrm{i}}

\newcommand{\h}[1]{\hat{#1}}

\usepackage{makecell}

\begin{document}

\author{Joel Steinegger${}^{\orcidlink{0009-0006-4845-1742}}$}
\email{joel.steinegger@dlr.de}
\affiliation{German Aerospace Center (DLR), 
Institute of Frontier Materials on Earth and in Space, Functional, Granular, and Composite Materials, 
51170 Cologne, Germany}

\author{Debasish Banerjee${}^{\orcidlink{0000-0003-0244-4337}}$}
\email{D.Banerjee@soton.ac.uk}
\affiliation{School of Physics and Astronomy, University of Southampton, University Road, SO17 1BJ, UK}

\author{Emilie Huffman${}^{\orcidlink{0000-0002-4417-338X}}$}
\email{ehuffman@wfu.edu}
\affiliation{Department of Physics and Center for Functional Materials, 
Wake Forest University, Winston-Salem, North Carolina 27109, USA}
\affiliation{Perimeter Institute for Theoretical Physics, Waterloo, Ontario N2L 2Y5, Canada}

\author{Lukas Rammelm\"{u}ller}
\affiliation{TNG Technology Consulting GmbH, Germany}

\title{Geometric fragmentation and anomalous thermalization in cubic dimer model}

\begin{abstract}
While quantum statistical mechanics triumphs in explaining many equilibrium phenomena, 
there is an increasing focus on going beyond conventional scenarios of thermalization. 
Traditionally examples of non-thermalizing systems are either integrable, or disordered. 
Recently, examples of translationally-invariant physical systems have been discovered
whose excited energies avoid thermalization either due to local constraints (whether exact 
or emergent), or due to higher-form symmetries. In this article, we extend these investigations 
for the case of 3D $U(1)$ quantum dimer models, which are lattice gauge theories with 
finite-dimensional local Hilbert spaces (also generically called quantum link models) with 
staggered charged static matter. Using a combination of analytical and numerical methods, 
we uncover a class of athermal states that arise in large winding sectors, when the system 
is subjected to external electric fields. The polarization of the dynamical fluxes in the 
direction of applied field traps excitations in 2D planes, while an interplay with the 
Gauss Law constraint in the perpendicular direction causes exotic athermal behaviour due 
to the emergence of new conserved quantities. This causes a geometric fragmentation of the 
system. We provide analytical arguments showing that the scaling of the number of fragments 
is exponential in the \emph{linear} system size, leading to \emph{weak fragmentation}. Further, 
we identify sectors which host \emph{fractonic} excitations with severe mobility restrictions. 
The unitary evolution of fragments dominated by \emph{fractons} is qualitatively different from
the one dominated by \emph{non-fractonic} excitations. 
\end{abstract}

\maketitle
\tableofcontents

\section{Introduction} \label{sec:intro}
  Interacting quantum many-body systems are expected to thermalize under unitary time evolution
when their initial states are simple Fock states. Consequently, time-evolved pure states obtained 
from different Fock states with the same energy density cannot be distinguished using purely local 
operators, even when they have different values of another local observable. To reconcile this 
phenomenon with the quantum mechanics of closed systems, dominated by the time-reversible 
Schr\"{o}dinger equation, one may invoke the eigenstate thermalization hypothesis (ETH) 
\cite{Deutsch1991,Srednicki1994,Rigol2008,Alessio2016}. 

 The ETH may be stated in terms of the matrix elements of observables in the eigenbasis of the 
 Hamiltonian as
\begin{equation} \label{eq:ETH}
    O_{mn} = O(\overline{E}) \delta_{mn} + e^{-S(E)/2} f_O (\overline{E},\omega) R_{mn},
\end{equation}
where $m,n$ are indices for the energy spectrum $E_n$, $\overline{E}=(E_m+E_n)/2,$ and 
$\omega = E_n - E_m$. Then $S(\overline{E})$ and $O(\overline{E})$ are the thermodynamic 
entropy and expectation value of the observable in the microcanonical ensemble at energy 
$\overline{E}$, respectively, 
$f_O$ is a smooth function of the arguments ($\overline{E},\omega$), and $R_{mn}$ are random real or 
complex variables with zero mean and unit variance \cite{Alessio2016}. Expectation values of 
local observables thus thermalize in $\mathcal{O}(1)$ time (in units of coupling) to the 
microcanonical ensemble due to the exponential damping offered by the entropy.

  Since most systems in nature thermalize, it is interesting to understand the 
conditions under which systems do not. Integrable and disordered systems offer some of the 
well-explored routes to violate ETH \cite{Tomasi2019,Alet2018}. Moreover, the initial 
experimental exploration of quantum dynamics on a Rydberg atom-based quantum simulator 
\cite{Bernien2017} led to the discovery of quantum many-body 
scars (QMBS) \cite{Turner2018a,Turner2018b,Choi2019}, which consist of a set of eigenstates with 
anomalously low entanglement entropy (and anomalous values of other observables) embedded in an 
otherwise ETH-satisfying spectrum. When starting from states with a high overlap with these 
QMBS, the unitary evolution may keep the system within the subspace of these scars, resulting in a
longer thermalization time. A theoretical description of this phenomena can be obtained by modelling
the system by the so-called PXP model \cite{Sachdev2002}, which has since been found 
\cite{Surace:2019dtp} to be identical to the spin-$\frac{1}{2}$ quantum-link Schwinger model 
\cite{Banerjee:2012pg,Huang:2018vss}. This has opened the door to large-scale quantum simulations 
of lattice gauge theories. Subsequently a plethora of non-ergodic behaviours have been discovered in 
strongly interacting systems \cite{Chandran:2022jtd,Pizzi:2024urq}. QMBS have been observed in a 
wide variety of physical systems from spin models  
\cite{Shiraishi2017,StarkMBL2019,Surace2021,Wildeboer2021,Mukherjee:2020lwa, Mukherjee_2020,
Serbyn:2020wys,Moudgalya:2021xlu,Moudgalya:2022nll,Udupa:2023rjg,Lerose:2023akj,Chandran:2022jtd,
Pizzi:2024urq,Pal_2025}, fermionic theories 
\cite{Pakrouski2020,Pakrouski2021,Moudgalya:2020eld,Schindler2022,Kolb2023}, 
and as well as in lattice gauge theories both with and without matter 
\cite{Banerjee:2020tgz,Biswas:2022env,Z2LGTscars,Halimeh:2022rwu,Desaules:2022ibp,Desaules:2022kse,
Hayata2023,Budde:2024rql,Osborne:2024zpx,Calajo:2024bvs}, and even in cases when gauge theories
are disordered \cite{Sau:2024uur}.

\begin{figure}[h!]
  \centering
  \includegraphics[scale=0.5]{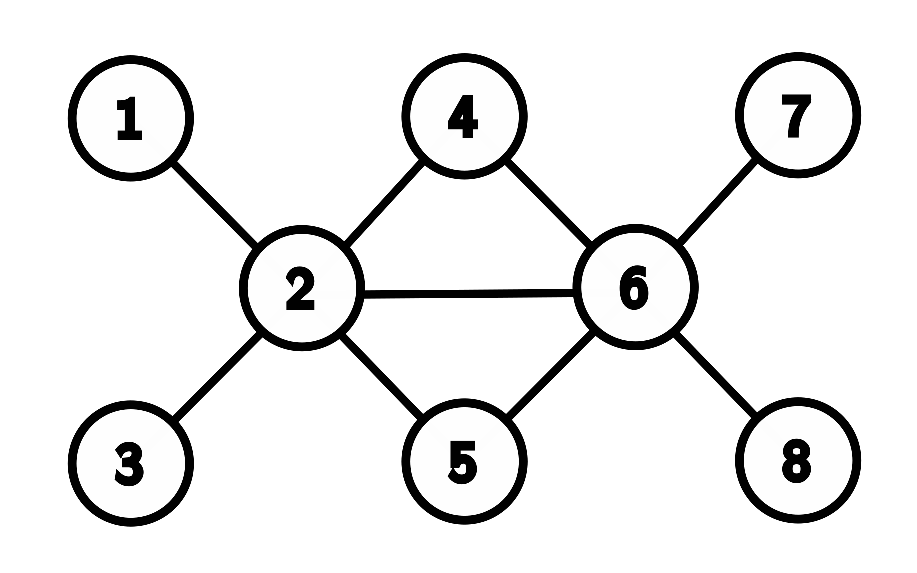}
  \hspace{1cm}
  \includegraphics[scale=0.5]{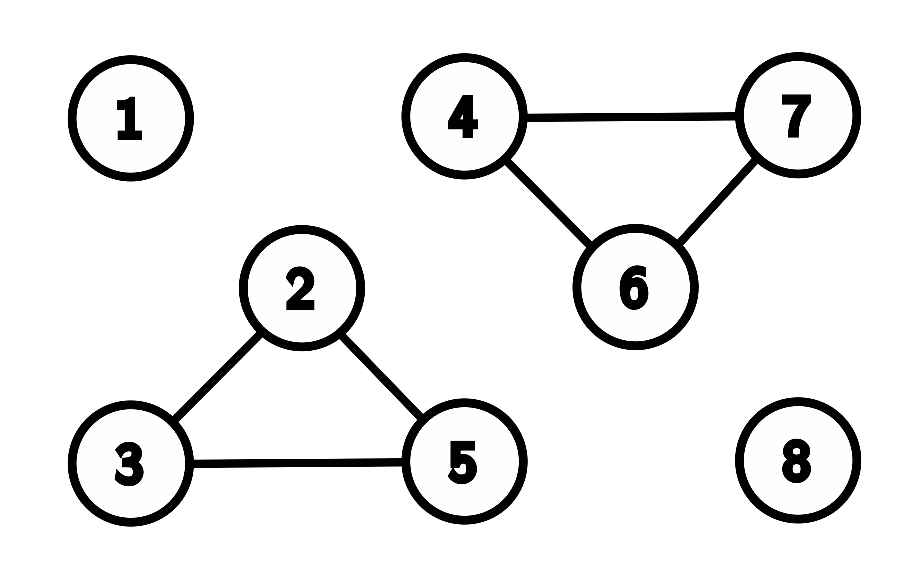}
  \caption{Consider an example of a quantum system with eight states which cannot be further 
  distinguished by local or global symmetries. The model is said to be {\bf ergodic} if the 
  Hamiltonian connects all the states (left). The same quantum system transforming under a
  different Hamiltonian which does not connect all the states, and the model is considered 
  to be {\bf fragmented} (right).}
  \label{fig:frag}
\end{figure}

 Another example of ETH-violating behaviour arises when the Hilbert space is fragmented into sectors 
that cannot be distinguished by global symmetries. Certain scenarios such as disorder-free 
localization 
\cite{Brenes:2017wzd,Smith:2017cpa,McClarty2020,Karpov2021,Osborne:2023wgd,Jeyaretnam:2024tkj} 
sometimes can be mapped to systems with local gauge or subsystem symmetries. Therefore anomalous 
thermalization in such systems can be understood as an incoherent sum of different (gauge) sectors, 
each thermalizing at its own pace. Other scenarios such as \emph{weak} and \emph{strong fragmentation} 
are more subtle, and can emerge without requiring local microscopic symmetries. 
To understand these scenarios, imagine that Fock states (in a suitable computational 
basis, or one which can be easily prepared in the lab) are represented as vertices of a graph, and 
the Hamiltonian represents the connections (bonds) of the graph. If the resulting graph is a single 
connected object, the system is ergodic, but if it instead is divided into disconnected sectors, the 
number of which grows with the system size, then the system can potentially evade thermalization
as postulated by the ETH. A common terminology used in this case is to say that the system is
\emph{fragmented}. The cartoon representation of \cref{fig:frag} for a Hamiltonian connecting 
different states of a finite system provides an intuition regarding fragmentation. 

A further distinction between 
fragmentation scenarios lies in whether one encounters a measure zero of ETH-violating states 
in the thermodynamic limit (\emph{weak fragmentation}), or whether the number of 
fragments grows exponentially with the volume but no single fragment 
dominates in the thermodynamic limit (\emph{strong fragmentation})
\cite{Sala2020,Khemani2020,Mukherjee2021,Moudgalya2021,Changlani2021,Chattopadhyay2023,
Neupert2022,Ghosh2024,Ciavarella:2025zqf,kwan2023minimalhubbardmodelsmaximal,
Harkema2024,Stahl2025}. 
The former case is similar to QMBS leading to eventual thermalization, but the latter can display
behaviour distinct from QMBS scenarios, in particular the dramatically reduced mobility
of excitations. A simple way to mathematically classify fragmentation is to consider the
scaling of the ratio $n/{\cal N}$, where $n$ is the number of states in the largest 
fragment (in a suitable computational basis) while ${\cal N}$ is the total number of
states (in the same basis) in the entire Hilbert space. If this ratio approaches an
${\cal O}(1)$ number in the thermodynamic limit, while exhibiting anomalous behaviour 
(e.g. QMBS states) at finite lattice sizes, the system is said to be \emph{weakly fragmented}, 
while if the ratio scales as ${\rm e}^{-a V}$ (where $V$ is the volume of the system) then 
the system is said to be \emph{strongly fragmented}. Concretely, this can be converted to a
comparison of the \emph{entropy density}, $s_{\rm frag}$, of the biggest fragment with 
the thermodynamic \emph{entropy density}, $s_{\rm tdyc}$. For \emph{weak fragmentation},
one has $s_{\rm tdyc} = s_{\rm frag}$ in the theromodynamic limit, while for the case of 
\emph{strong fragmentation}, $s_{\rm tdyc}$ and $s_{\rm frag}$ differ by $a \sim {\cal O}(1)$.

 In this article, we discuss a form of \emph{geometric fragmentation} arising in a class of 
quantum many-body systems in three spatial dimensions which have local conservation laws, better
known as quantum link gauge theories. Specific examples of these models, such as quantum dimer 
models, have been extensively used in the quantum condensed matter community to discuss the 
non-magnetic phases of electrons at low temperatures where electrons can form singlets with 
their nearest neighbours, and are relevant as microscopic models of high-temperature 
superconductors \cite{Rokhsar1988,Kivelson1987,Moessner2003,Bonca2007}. Realized on a lattice 
structure consisting of corner-sharing tetrahedra (pyrochlores), such models are relevant in 
the physics of spin-ice compounds and spin-liquid phases in those systems
\cite{Hermele2004,Banerjee2008,Sikora2009,Sikora2011,Gingras:2013xoa,Benton2012,Ross:2011zz,
Castelnovo:2012cda,Shah2025}. In the context of particle physics, these quantum link models 
\cite{Horn1981,Orland1990,Chandrasekharan1997,Brower1999,Liu:2021tef,Berenstein:2022wlm} 
were proposed as generalizations of Wilson's lattice gauge theory \cite{Wilson1974,Kogut1975} 
to develop better classical algorithms for quantum chromodynamics (QCD) 
\cite{Banerjee2013,Banerjee2014,Tschirsich2019,Magnifico2021,Banerjee:2021zed,Chandrasekharan:2025smw}. 
Thanks to recent experimental developments on quantum simulations and computations, these
models are especially suited to be realized on near-term devices in order to study phenomena 
of interest in particle physics, particularly for problems which defy classical simulation techniques
\cite{Banerjee:2012pg,Gonzalez2017,Bender2018,Ott2021,Banuls2020,Halimeh:2023lid,Bauer:2023qgm}. 

 Most of the studies so far have concentrated on the phase diagram at zero temperature and at finite 
temperatures, which is natural given the available classical methods, as well as the experimentally 
relevant physics. However, thanks to the new tools for quantum simulation, there is an increasing 
interest in the conditions leading to (lack of) thermalization in these models. We show that in the 
presence of (large) external electric fields, magnetic excitations get trapped in two-dimensional 
planes hindering thermalization. We are able to use analytical techniques to characterize certain 
aspects of the anomalous states, while for other cases we display the athermal behaviour numerically. 
Our starting point is the quantum dimer model on the cubic lattice. We then confine 
ourselves to a particular sector of the model, characterized by the largest winding number in 
a direction, which displays fragmentation. 

The rest of the paper is organized as follows. In \cref{sec:model}, we introduce the model and 
discuss the constraints imposed by the three-dimensional Gauss law, as well as the global winding 
number symmetry that plays a crucial role in inducing geometric fragmentation. We then describe 
the observables used to characterize the anomalous dynamics. The concept of geometric fragmentation 
is introduced in \cref{sec:geoFrag}, with both its origin and nature examined in detail in
\cref{sec:fragmentation_doped_lattice}. In \cref{sec:real_time_dynamics}, we explore the emergence 
of fractons within certain fragmented subspaces and how they give rise to athermal dynamics — both in 
fragments dominated by fractons and in those that are not. Finally, in \cref{sec:analytical_solution}, 
we present an analytical solution for the eigenstates and eigenenergies of the fractonic fragments. 
We conclude by summarizing our main findings and providing an outlook in \cref{sec:outlook}.

\section{Model and Observables} \label{sec:model}
\subsection{Doped bosonic Quantum link model}
We begin by introducing the bosonic $U(1)$ Abelian Quantum link model (the fermionic version 
was introduced in \cite{Huffman2022}). Although we will focus on the Hilbert space structure 
and the real-time dynamics of the model in three dimensions, an understanding of the model 
in two dimensions equips the reader with the insight to appreciate the special dynamics we 
show later. The model is defined as follows on square and cubic lattices, 
\begin{equation} \label{eq:Hamiltonian_1}
H = \frac{g^{2}}{2}\sum_{x, \h{i}} E^{2}_{x, \h{i}} - 
J \sum_{\square} (U_{\square}+U_{\square}^{\dagger})+\lambda \sum_{\square}(U_{\square}+U_{\square}^{\dagger})^{2},
~~~~U_{\square} = U_{x, \h{i}} U_{x+\h{i},\h{j}} U^{\dagger}_{x+\h{j},\h{i}} U^{\dagger}_{x,\h{j}}.
\end{equation}
The degrees of freedom of the model are defined on links joining two adjacent lattice sites, and
labelled with the subscript $(x,\h{i})$, with $x$ as a site and $\h{i}$ as a unit vector in 
a spatial direction. The first term in \cref{eq:Hamiltonian_1} is the electric field energy 
(square of the electric fluxes, $E_{x,i}$), while the second term is the magnetic field 
energy, and the third term is the Rokhsar-Kivelson (RK) term. The latter two terms are expressed 
via plaquette operators, $\mathrm{U}_{\square}$. We will choose a computational basis which 
is diagonal in the electric fluxes, and thus the term linear in plaquette operators is 
off-diagonal in this basis and will be the kinetic operator 
$H_{\rm kin} = \sum_{\square} (U_{\square}+U_{\square}^{\dagger})$ while the term 
quadratic in the plaquette operator is diagonal in the flux basis is 
$H_{\rm pot} = \sum_{\square}(U_{\square}+U_{\square}^{\dagger})^{2}$. 

The speciality of the quantum link formulation is that these operators can be represented by
a finite-dimensional Hilbert space, and are characterized by the representations of $SU(2)$
algebra. In particular, the operators satisfy the following commutation relations: 
\begin{align} \label{eq:commutation}
[E_{x,\h{i}}, U_{y,\h{j}}] = U_{x,\h{i}} \delta_{x,y} \delta_{i,j};~~~
[E_{x,\h{i}}, U^\dagger_{y,\h{j}}] = -U^\dagger_{x,\h{i}}\delta_{x,y} \delta_{i,j};~~~
[U_{x,\mu}, U^\dagger_{y,\nu}] = 2E_{x,\mu} \delta_{x,y} \delta_{i,j}.
\end{align}
The Hamiltonian has a local $U(1)$ symmetry generated by the local lattice Gauss law operator:
\begin{equation} \label{eq:gauss_operator}
G_{x}=\sum_{i}(E_{x,\h{i}}-E_{x-\h{i}, \h{i}});
~~~[G_{x}, H] = 0~~\textrm{for all}~~x.
\end{equation}
This causes the many-body Hilbert space to break into exponentially many sectors, labelled by 
quantum numbers of the local charge, $G_{x}$. A typical choice in particle physics is to 
define physical states as being annihilated by the Gauss law: $G_{x} \ket{\Psi}=0$, which 
corresponds to zero charge. In contrast, physical states of a quantum dimer model (QDM) have doped 
(immobile) staggered charges, mathematically represented as $G_{x}\ket{\chi}= Q (-1)^{x}\ket{\chi}$, 
where $(-1)^{x}$ is the site parity, and $Q$ is the quanta of charge. 

\begin{figure}
    \centering
    \includegraphics[scale=0.3]{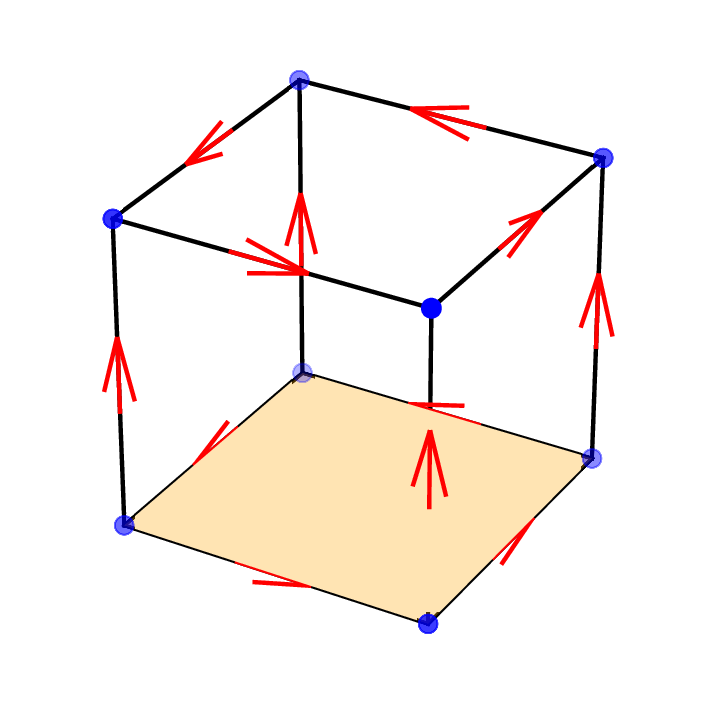}
    \includegraphics[scale=0.3]{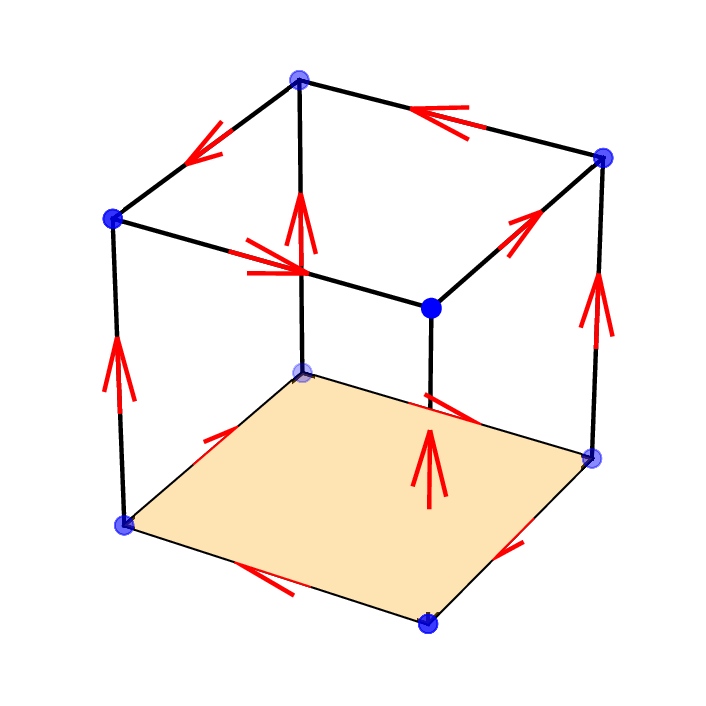}
    \includegraphics[scale=0.3]{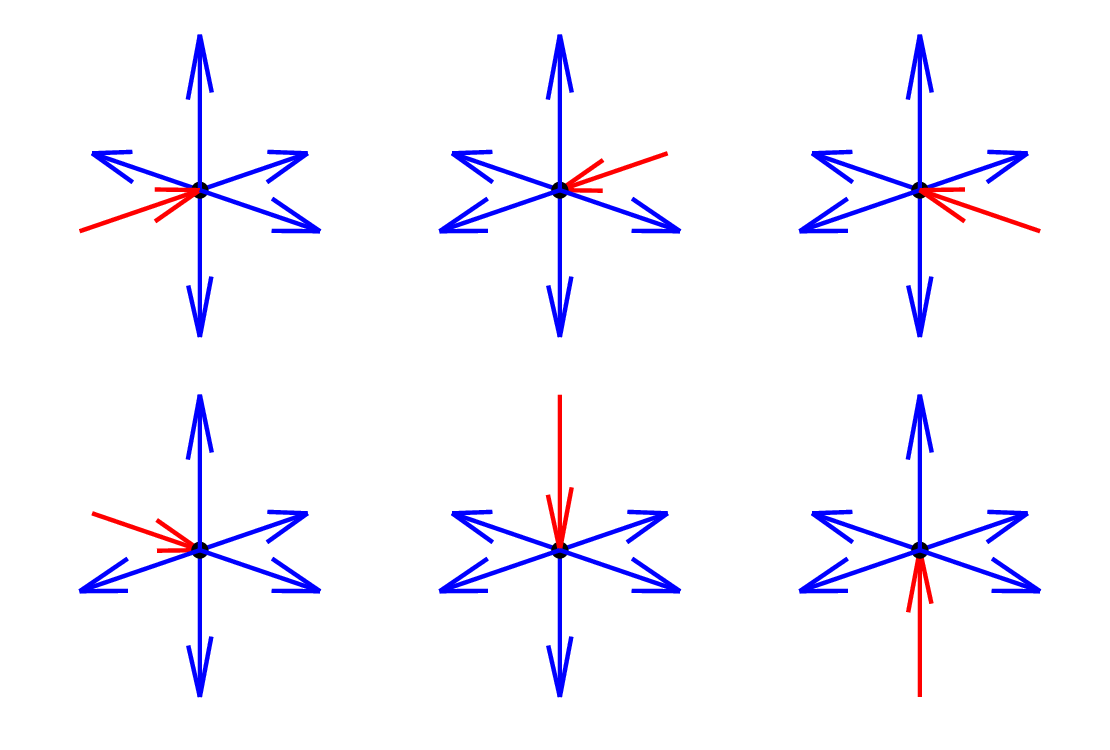}
    \caption{ An elementary cube whose top and bottom faces are flippable in the anti-clockwise 
    fashion (left). The $J$-term of the Hamiltonian acts on the lower (shaded) face and converts 
    it into a clockwise flippable plaquette (middle). Hamiltonian flips preserve the local charge. 
    (Right) States which are allowed for $Q=2$ condition.}
    \label{fig:latGeo}
\end{figure}

Any representation of the operators fulfilling Eq. \ref{eq:commutation} is allowed to define
the theory. The well-known Wilson-type lattice gauge theory uses the quantum rotor as its degree of 
freedom, generating an infinite-dimensional representation on each of the links \cite{Kogut1975}. 
A finite-dimensional representation is obtained by using spin-$S$ operators $\vec{S}_{x,\h{i}}$ 
as degrees of freedom on the link, with the following identifications:

\begin{equation} \label{spin_S_representation}
E_{x,\h{i}}=S^{3}_{x,\h{i}};~~~ U_{x,\h{i}}=S^{+}_{x,\h{i}};~~~ U^{\dagger}_{x,\h{i}}=S^{-}_{x,\h{i}}.
\end{equation}
In the limit of large $S$, the Wilson lattice gauge theory is approached \cite{Zache2021}. 
In this work, we 
consider the other extreme limit of $S=\frac{1}{2}$ with a two-dimensional local Hilbert space,
and work in the electric flux basis. The electric field energy is constant in this representation
and can be ignored. Working in the electric field basis, the states can be represented by flux 
arrows pointing outwards (inwards) to the site representing flux $\frac{1}{2}$ ($-\frac{1}{2}$).
A pictorial representation of the action of the Hamiltonian on an individual state is given in 
\cref{fig:latGeo}. Moreover, physical states can be locally classified using the Gauss Law. The 
number of states 
allowed decreases with the increase in $Q$: for three-dimensional cubic lattices, $Q=0$ admits 
20 states \cite{Huffman2022}, $Q=1$ allows 15 states, $Q=2$ allows 6 states, and $Q=3$ allows 
only a single state. \cref{fig:latGeo} shows the case of the local charge $Q=+2$. We do not draw all
the states for the other Gauss Law conditions, but the reader is encouraged to draw them for
their understanding.

\subsection{Winding number symmetry}
 The Hamiltonian of the system is invariant under (global) point group symmetries (rotations,
 reflections, and translations), as well as charge conjugation. Our work focusses on the remaining 
 continuous global center symmetry, which gives rise to conserved flux winding number sectors.
 On an $L_x \times L_y \times L_z = V$ (three-dimensional) lattice with periodic boundary conditions, 
 there are three separately conserved winding numbers $W=(W^{3D}_{x}, W^{3D}_{y}, W^{3D}_{z})$:
\begin{equation} \label{Wx_Wy_Wz}
 W^{3D}_{x} = \frac{1}{L_{y}L_{z}} \sum_{r} E_{r,x},~~W^{3D}_{y} = \frac{1}{L_{x}L_{z}} \sum_{r} E_{r,y},~~
 W^{3D}_{z} = \frac{1}{L_{x}L_{y}} \sum_{r} E_{r,z}.
\end{equation}
Illustrations of the winding number calculations in 3D are shown in \cref{fig:winding_calculation_3D}. 
$W_{i}$ ranges in integer steps from $-L_{\i}/2$ to  $L_{i}/2$. Since we consider 
the physics for the largest winding sector in a given ($z$) direction, we will also need the corresponding two dimensional winding numbers, $W=(W^{2D}_{x}$, $W^{2D}_{y})$:
\begin{equation} \label{Wx_Wy}
 W^{2D}_{x} = \frac{1}{L_{y}} \sum_{r} E_{r,x},~~~W^{2D}_{y} = \frac{1}{L_{x}} \sum_{r} E_{r,y}.
\end{equation}
The corresponding calculation for $W^{2D}_i$ is illustrated in \cref{fig:winding_calculation_2D}. \\
\noindent
\begin{minipage}[t]{0.49\textwidth}
  \centering
  \includegraphics[width=0.8\textwidth]{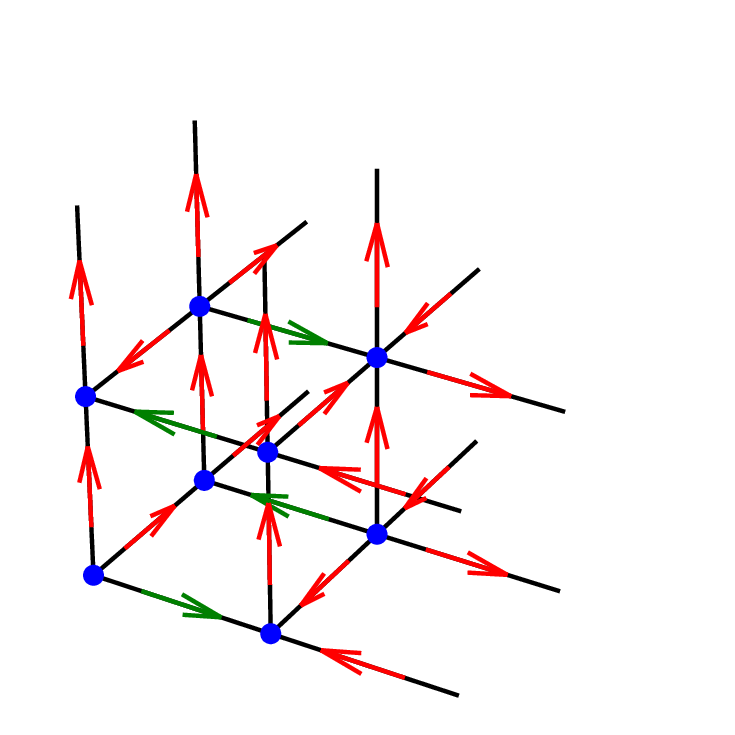}
  \captionof{figure}{\label{fig:winding_calculation_3D} An example state which maximal winding 
  in the $z$-direction: $(W_x,W_y,W_z)=(0,0,2)$. The quantity $W^{3D}_{x}$ is obtained by 
  summing the flux contributions highlighted in green.}
\end{minipage}
\hfill
\begin{minipage}[t]{0.49\textwidth}
  \centering
   \includegraphics[width=0.8\textwidth]{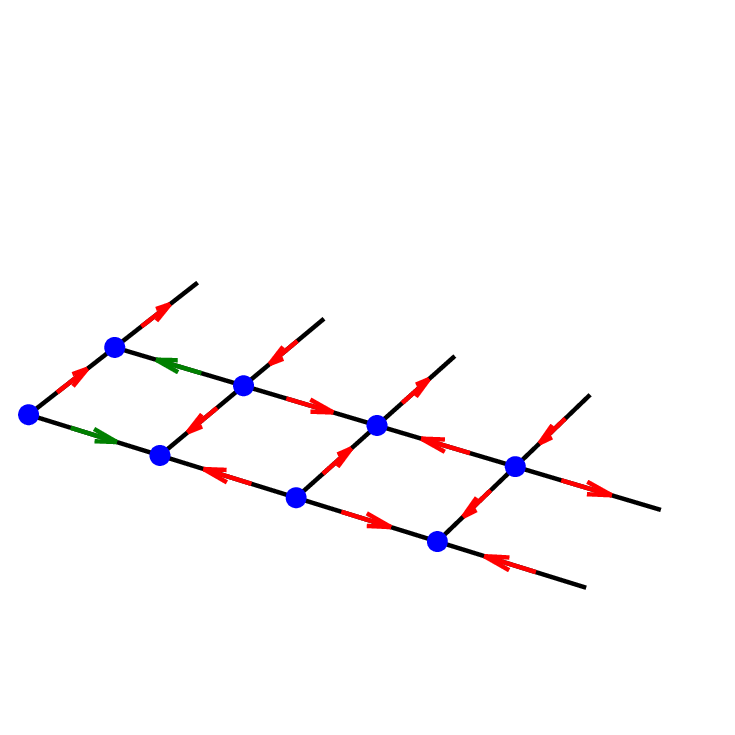}
  \captionof{figure}{\label{fig:winding_calculation_2D} Calculation of the winding number 
  in two-dimensional planes. In this example, we have the $2 \times 4$ lattice with the
  winding number $(W^{2D}_x,W^{2D}_y)=(0,0)$. The quantity $W^{2D}_{x}$ is obtained 
  by summing the flux contributions highlighted in green.
  }
\end{minipage}

\subsection{Time dependent observables}
 Before discussing the physics of how disconnected sectors arise in 
these models, let us briefly mention the observables we use to track the consequences of
ergodicity breaking. Since thermalization implies the loss of memory of initial conditions,
a standard technique is the study of real-time dynamics of the expectation values of operators, 
fidelities, and measures of entanglement starting from initial states which can be 
\emph{easily} prepared both theoretically and experimentally. In \cref{sec:real_time_dynamics}, 
we will examine the unitary evolution of the system, comparing behaviours in different subspaces,
and contrast the sectors which thermalize from those which do not. To exemplify how fragmentation 
leads to ETH-breaking in this model and to illustrate our analytical findings about atypical
dynamics of the model we numerically calculate the following time-dependent measures: 
the fidelity
\begin{equation}\label{eq:fidelity}
F(t)=|\braket{\Psi(0)|\Psi(t)}|^2,
\end{equation}
which measures the overlap with the initial state. Thermalization implies that $F(t)$ goes to
zero very rapidly, and typically never becomes an $\mathcal{O}(1)$ quantity. When this quantity regularly
increases, some form of ergodicity breaking is suspect. The Shannon entropy, defined as
\begin{equation} \label{shannon_entropy}
S(t)= -\sum_{n} |c_{n}|^{2} \text{log}_{2}(|c_{n}|^{2}) \text{ with } \ket{\Psi(t)}=\sum_{n}c_{n}\ket{n},
\end{equation}
measures the spread of the wavefunction in the Fock space. Note that the numerical value
is dependent on the basis chosen to label the Hilbert space (which we denote as $\ket{n}$ in a
particular winding flux sector), but other (local) basis choices are still expected to show the
difference between thermalizing and non-thermalizing behaviours. When started in a pure state, it 
is zero, but will increase to an $\sim \mathcal{O}(1)$ quantity if the wavefunction has support 
in the entire Fock space. The maximally entangled state scales as the logarithm of the total 
Hilbert space. Finally, we also track the evolution of kinetic and potential energy, which we 
label as
\begin{equation} \label{kinpot_energy}
E_{\rm kin}(t)=\bra{\Psi(t)} H_{\rm kin} \ket{\Psi(t)}, \qquad 
E_{\rm pot}(t)=\bra{\Psi(t)} H_{\rm pot} \ket{\Psi(t)},
\end{equation}

 In all our numerical examples, we start the state at $t=0$ in one of the flux basis states of 
the investigated subspace: $\ket{\Psi(0)}=\ket{n}$, and use $\Delta t = 0.001$ as the time step 
size uniformly in all our calculations. However, since we are always using exact methods for the 
calculation, the value at any arbitrary time can be computed. 

\section{Geometric Fragmentation} \label{sec:geoFrag}
 In the absence of any staggered static charges, i.e., in the sector characterized by $G_x = 0$, 
the model is expected to ergodic for zero winding number sectors. While we are aware that this 
is the case in $d=2$ \cite{Biswas:2022env}, a similar check does not (yet) exist in $d=3$ to 
the best of our knowledge.  As the system is 
doped using staggered static charges, the allowed Gauss law realizations further constrain the
set of allowed states, and risk an otherwise connected Hamiltonian graph to fragment. 
This is even more so if the system is subjected to an external field which locks the 
electric fluxes in the direction of the field, but keeps those transverse to the field
unaffected. In this section, we consider a particular limiting case of the quantum dimer model 
(i.e. where the constraint $G_x = (-1)^x$ selects the physical states, and allows for 15 states
locally on each site), with external electric field. This is equivalent to considering the model 
in the highest winding number sector in the direction of the field, for example in the sector 
$(0,0,W_z^{\rm max})$ without any loss of generality. An example state in this sector is shown
in \cref{fig:winding_calculation_3D}. Let us now discuss how fragmentation may arise in this
scenario. 

\subsection{\label{sec:fragmentation_doped_lattice}Fragmentation in the doped QLM}
 In the maximal winding number sector of the dimer model, all the $z$-fluxes (in the direction 
of the external electric field) point in the same direction. This holds irrespective of the lattice
size. In particular, this means that the spatial plaquettes in the $xz$-plane and the $yz$-plane cannot
be made flippable since the $z$-links in each of these plaquettes always point in the same direction. 
Flippable plaquettes can only exist in the $xy$-plane. This leads to the key result that
the maximal flux states can be represented as $L_{z}$ stacked 2D quantum dimer models (QDMs), 
which stagger the charge as one moves in the $z$-direction at a fixed value of $(x,y)$

This means that we can effectively describe the basis states in the flux basis of the 
$(0,0,W^{\mathrm max}_z)$ sector as:
\begin{equation}
\label{eq:stacked_2D_lattice state}
\ket{\Psi_{\textrm 3D}} = \ket{\Psi^{1}_{\textrm 2D}} \otimes \ket{\Psi^{2}_{\textrm 2D}} \otimes \hdots 
 \ket{\Psi^{L_z}_{\textrm 2D}},
\end{equation}
since the fluxes in the $z$-direction are fixed. The effective Hamiltonian reduces to a 
sum of operators only acting on the $xy$ planes:
\begin{equation}
\label{eq:amiltonian_on_stacked_2D}
 H = H^{2D}_{1} \otimes \mathds{1}^{\otimes (L_z -1)} 
   + \mathds{1}\otimes H^{2D}_{2}\otimes \mathds{1}^{\otimes (L_z -2)} + \hdots 
   + \mathds{1}^{\otimes (L_z -1)} \otimes H^{2D}_{L_{z}}.
\end{equation}
The Hamiltonian can be taken to be $\mathds{1}$ for the terms in the $xz$- and $yz$-planes since their plaquettes are
not flippable by construction. Furthermore, the effective 2D QDMs must be stacked in a
pattern such that the correct winding sectors $W^{3D}_{x}$ and $W^{3D}_{y}$ are reproduced.
This is responsible for creating a further constraint in the perpendicular directions 
even though the external field does not directly affect $E_x$ and $E_y$ links. 
The winding number for a direction $i$, $W^{3D}_i$, is obtained by adding the $i$-th flux 
through a plane perpendicular to $i$. On the other hand, $W^{2D}_i$, is obtained by summing 
the $E_i$ along a line perpendicular to the links. Therefore, we can calculate the winding 
number of $\ket{\Psi_{\textrm 3D}}$ by summing the winding numbers of $L_z$ stacked 2D QDMs: 
\begin{equation}
\label{3D_W}
  W^{3D}_x=\sum_{i=1}^{L_{z}} W^{2D}_{x,i},\qquad
  W^{3D}_y=\sum_{i=1}^{L_{z}} W^{2D}_{y,i} 
\end{equation}

 In the maximal winding sector $(W_x, W_y, W_z^{\rm max})$, this provides a way to identify
the new conserved quantity, $W^{2D}_i$, since $W^{2D}_i$ of the 2D planes cannot change by 
applying the Hamiltonian. The only flippable plaquettes are in the $xy$-planes which cannot  
communicate via frozen $E_z$ fluxes on the $xz$- and the $yz$- planes. 
It is important to realize this is not the case for states in arbitrary winding number 
states in 3D and is not due to a global symmetry of the Hamiltonian. Particularly, if we 
can fulfill \cref{3D_W} with different combinations of 2D QDM planes, they would not be 
connected by the Hamiltonian. This results in geometric fragmentation of the Hilbert space. 
Naively, one may think that this always implies fragmentation for maximal flux in one 
direction, but there is no guarantee that there exist at least two ways to fulfill 
\cref{3D_W}. To predict fragmentation definitively, we must also obtain winding subspaces 
of an arbitrary lattice of size $L_x \times L_y \times L_z$ consist of 
states that fulfill \cref{3D_W} in at least two different ways. In \cref{sec:fragntation_proof}, 
we analytically prove this to be the case for a large class of examples. 

\begin{figure}[h!]
  \centering
     \includegraphics[width=0.8\textwidth]{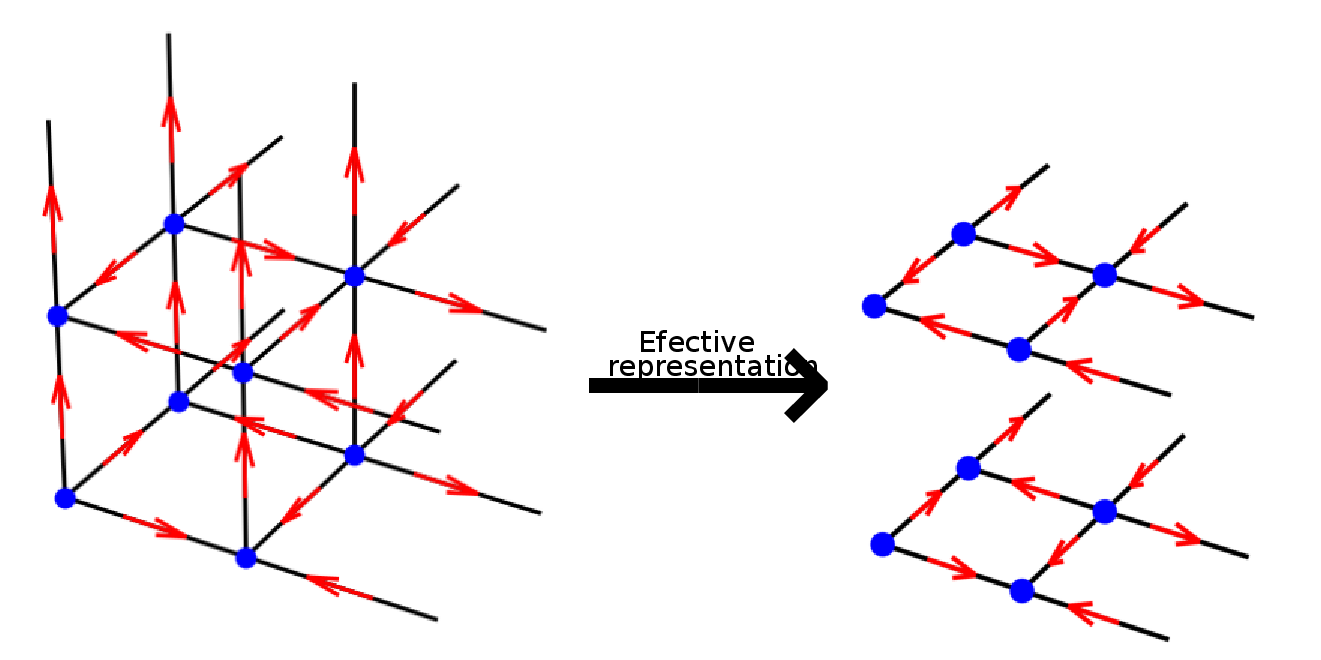}
\caption{\label{fig:effective_representation} When the winding numbers are maximal in a
given direction, it is possible to represent the states as 2D planes since the dynamics
is frozen in the perpendicular direction of the plane.}
\end{figure}

In the main text, let us instead show how this arises by simple examples. Consider the 
$4\times 2\times 2$ lattice in the winding sector $W^{3D}=(1,1,4)$: this is fragmented 
into two subspaces, each consisting of eight states. \cref{table:4x2} shows the 
subspaces of the QDM alongwith the number of states in each. The geometry makes the
system be treated as two parallel alternating dimer planes of size $2\times 4$, which 
have to fulfill \cref{3D_W}. An example of this \emph{effective representation} for 
the $2\times2\times2$ lattice is shown in \cref{fig:effective_representation}. To 
fulfill these equations with states 
depicted in \cref{table:4x2}, there are only two ways: 
\begin{enumerate}
\item $W^{2D}_{x}=1$ and $W^{2D}_{y}=0$ for the first plane, 
      and $W^{2D}_{x}=0$ $W^{2D}_{y}=1$ for the second plane
\item $W^{2D}_{x}=0$ and $W^{2D}_{y}=1$ for the first plane, 
      and $W^{2D}_{x}=1$ $W^{2D}_{y}=0$ for the second plane
\end{enumerate}
 From here we see that that the sector is fragmented into two subspaces. In both solutions, 
one plane contains eight possible states, and the other plane contains one possible state, 
resulting in the fact that both subspaces have eight states as predicted by a brute-force
numerical calculation, where we construct the Hamiltonian graph for the Fock space and 
check the number of connected components. Therefore, the numerical example matches our 
maximal-flux description of how fragmentation occurs. Similarly, in the sector $(1,4,1)$,
one finds a similar fragmentation, as is of course predicted by symmetry. 
It is instructive to consider the winding sector 
\( W^{3D} = (3,0,2) \) of the \( 2 \times 2 \times 4 \) lattice as a second example. This winding 
subspace is fragmented into four distinct sectors, each containing four 
3D quantum dimer model (QDM) states. Due to the geometry of the system, 
each state can be interpreted as comprised of four parallel, alternating 
dimer planes of size \( 2 \times 2 \), each subject to the constraints 
imposed by \cref{3D_W}. There exist four distinct ways to satisfy these 
constraints using distinct \( 2 \times 2 \) subspaces of the 2D QDM. The 
possible subspaces of the \( 2 \times 2 \) QDM are shown in \cref{table:2x2}. 
The construction begins by selecting one of the four planes to host a 
configuration with winding numbers \( W^{2D}_x = 0 \) and \( W^{2D}_y = 0 \), 
while assigning \( W^{2D}_x = 1 \) and \( W^{2D}_y = 0 \) to the remaining 
planes. Since there are four possible choices for the plane in the 
\( W^{2D}_x = 0 \), \( W^{2D}_y = 0 \) configuration, this results in four 
distinct fragments, each containing four states, within the subspace.\\

\begin{table}[htbp]
  \centering
  \begin{subtable}[t]{0.45\textwidth}
    \centering
    \resizebox{\textwidth}{!}{
      \begin{tabular}{|c|c|c|c|}
        \hline
        $W_x$ & $W_y$ & \# States & Fractons \\
        \hline
         2    &  0    &    1      & \ding{55} \\
        -1    &  0    &    8      & \checkmark \\ 
         0    & -1    &    1      & \ding{55} \\
         0    &  0    &   16      & \ding{55} \\
         0    &  1    &    1      & \ding{55} \\
         1    &  0    &    8      & \checkmark \\
         2    &  0    &    1      & \ding{55} \\
        \hline
      \end{tabular}
    }
    \caption{Winding subspaces of a $2 \times 4$ QDM.}
    \label{table:4x2}
  \end{subtable}
  \hfill
    \begin{subtable}[t]{0.45\textwidth}
    \centering
    \resizebox{\textwidth}{!}{
      \begin{tabular}{|c|c|c|c|}
        \hline
        $W_x$ & $W_y$ & \# States & Fractons \\
        \hline
         -1   &  0    &     1     & \ding{55} \\
          0   & -1    &     1     & \ding{55} \\
          0   &  0    &     4     & \checkmark \\
          0   &  1    &     1     & \ding{55} \\
          1   &  0    &     1     & \ding{55} \\
        \hline
      \end{tabular}
    }
    \caption{Winding subspaces of a $2 \times 2$ QDM.}
    \label{table:2x2}
  \end{subtable}
  \caption{Comparison of winding subspaces and fracton presence in $2 \times 2$ and 
  $2 \times 4$ lattices. Fractons are plaquette excitations which have reduced
  mobility along subdimensional manifold than the one in which the hamiltonian is
  defined.}
\end{table}

  Next we analyze the type of fragmentation --- strong or weak --- that emerges 
due to geometric constraints induced by maximal flux along a single direction in a 3D QDM. 
To quantify the nature of this fragmentation, we consider the 
following ratio in the thermodynamic limit ($L\to\infty$) of the cubic lattice $L\times L \times L$:
\begin{equation}
\label{eq:strong_frag_scaling}
  \mathcal{F} = \lim_{L\to\infty} \frac{N_{\mathrm{largest}}}{N_{\mathrm{total}}}
\end{equation}
where \( N_{\mathrm{largest}} \) denotes the number of states in the largest 
fragment within a given fragmented winding sector, and \( N_{\mathrm{total}} \) 
is the total number of states in that winding sector. If this quantity scales 
as \( \exp(-V) \), the system exhibits strong fragmentation, otherwise we will
claim weak fragmentation. 

 In the main text of this work, we concentrate on the winding sector \((0,0,W^{\rm max}_z)\) 
specifically to characterize the nature of its fragmentation. Based on the construction of 
stacking \( L_z \) 2D lattice planes along the \( z \)-direction, it follows directly that 
the largest fragment within this winding sector corresponds to stacking 
\( L_z \) copies of the \( (W^{2D}_{x}, W^{2D}_{y}) = (0,0) \) sector on top of one 
another in a lattice of size \( L_x \times L_y \times L_z \). Let's say the 
number of states in the 2D $(W^{2D}_{x}, W^{2D}_{y}) = (0,0)$ sector is $n_{0}$. It 
follows that $N_{\rm largest}=n_{0}^{L_{z}}$.  Our approach involves establishing 
upper and lower bounds on \(\mathcal{F}\) (\cref{eq:strong_frag_scaling}). The key 
problem with evaluating the scaling of \(\mathcal{F}\) 
is that $N_{\rm total}$ (total number of states across all fragments) is not easily 
computable since we are dealing with a constrained model in the thermodynamic limit. 
Thus we will instead find the scalings of two numbers: $N_{\rm min}$ which is a known subset of the total number of states and thus always smaller than $N_{\rm total}$, and $N_{\rm max}$, which is by construction larger than 
$N_{\rm total}$ for all $L$. We thus will have 
$N_{\rm min} < N_{\rm total} < N_{\rm max}$ for all $L$. 

  For $N_{\rm min}$, we note that in the thermodynamic limit, 2D lattice planes in the
following five winding number sectors $(0,0)$, $(0,\pm 1)$ and $(\pm 1,0)$ have the
same number of Fock states. To satisfy $W^{3D}_x=0$, we can combine $L_z/2$ pairs of 2D 
planes, such that \emph{both} elements of a single pair have $(0,0)$ windings, or 
one plane has $(1,0)$ and the other $(-1,0)$ winding (similarly for y-windings). 
Further, $n_1 \approx n_0$, where $n_1$ is the number of states in the $(1,0)$ winding 
sector (similarly for the y-winding). Thus, $N_{\rm min} \approx n_0^{L_z} 5^\frac{L_z}{2}$, 
(the solutions along the x- and y-directions are independent) and we can form an 
\emph{upper} bound by only considering these subset of allowed solutions to the 
winding constraints: 
\begin{equation} \label{eq:upper}
  \lim_{L\to\infty} \frac{N_{\rm largest}}{N_{\rm total}} < \mathcal{G}_{\rm upper}
   = \lim_{L\to\infty} \frac{N_{\rm largest}}{N_{\rm min}}=\lim_{L\to\infty} e^{-L\frac{\ln(5)}{2}}.
\end{equation}
On the other hand, we can obtain an $N_{\rm max}$ by overcounting the number of allowed solutions,
by including all the winding sectors (there are $L+1$ of them in each direction) and assuming 
that each 2D winding sector has the same number of states $\approx n_0$, when in reality the 
number of states in the sector decrease with the winding of the sector. Thus, 
$N_{\rm max} \approx n_0^{L_z} (L + 1)^{2 \frac{L_z}{2}}$, where $L = L_z$ and $L_z$ has been 
written out to show the pairing. This yields a lower bound $\mathcal{G}_{\rm lower}$:
\begin{equation} \label{eq:upper}
  \lim_{L\to\infty} \frac{N_{\rm largest}}{N_{\rm total}} > \mathcal{G}_{\rm lower}
   = \lim_{L\to\infty} \frac{N_{\rm largest}}{N_{\rm max}}=\lim_{L\to\infty}e^{-L\ln(L+1)}.
\end{equation}

 Evaluating the entropy density (where the total entropy is logarithm of the number of states), 
we find that the leading correction to the value in the thermodynamic limit is ${\cal O} (1/L^2)$ 
in the former case and ${\cal O} ( \ln(L)/L^2)$ in the latter case.
Since the leading correction vanishes as $L \to \infty$, the entropy density of the biggest fragment
is the same as that of the full theory, and thus we classify this scenario as \emph{weak 
fragmentation}. The actual behaviour of the leading correction in our model is in between these
limits, but will also go as ${\cal O} ( \alpha /L^2)$, where $\alpha$ has a weaker dependence
on $L$ than a logarithm. 

We emphasize that this leading behaviour is different from that of models which are
expected to thermalize according to the ETH. For example, the spin-$1/2$ antiferromagnetic 
Heisenberg model on the cubic lattice only has the total magnetization as the global conserved
quantity. If one considers the ratio of the entropy density in the largest magnetization sector
(which has $L^3/2$ spins with $S_z = 1/2$ and the other $L^3/2$ spins with $S_z = -1/2$), the leading
correction decays as ${\cal O} (1/L^3)$, \textit{faster} than the case of the 3D quantum dimer model
we have presented. Intuitively, the weak fragmentation in our case is a consequence of a genuine 
3D system fragmenting into individual 2D planes, with the excitations only confined within 
the 2D planes. 

  It can be shown that this plane pairing argument is enough to predict the exact leading correction 
to the thermodynamic limit. More precise upper and lower bonds could be made by establishing a 
scaling behaviour of $\frac{n_f (W)}{n_0}$ where $n_f$ is the number of states in an arbitrary 
winding $W$ sector of 2D $L\times L$ lattice, but we consider this unnecessary for our present 
purpose. Furthermore, similar arguments can be made for other winding sectors of this model if the 
amount of flux in one direction is maximal ($\lvert W^{3D}_{i}\rvert=L_{j} \cdot L_k/2$). We 
find that these sectors are also weakly fragmented. The proof of the weak fragmentation 
is provided in the \cref{sec:non_zero_winding_strong_fragmentation}.  

\subsection{\label{sec:real_time_dynamics}Real time dynamics and Thermalization}
 In \cref{sec:intro}, we pointed out that unitary dynamics provides another way to 
diagnose the presence of ETH-violating behaviour --- geometric fragmentation
in our case. The system initialized in one of the fragmented subspaces would 
thermalize slowly (or not at all in certain cases), requiring times much greater 
than $\mathcal{O}(1)$ time than expected if fragmentation were absent. In some cases, this 
ETH-violating behaviour can be related to the existance of quasiparticles, called 
fractons, with restricted mobility. We start this subsection by developing an 
analytical understanding of fractons in certain fragmented spaces. Using the 
(non-)existance of fractons, we can characterize the real-time dynamics 
for the 3D QDM with maximal flux in a given direction into two different 
categories. We contrast the dynamics for a fragmented sector of the model to 
another sector which thermalizes according to the prediction of ETH. 

\subsubsection{Fragmented subspaces}
 In \cref{sec:fragmentation_doped_lattice}, we established fragmentation in the 
maximal winding sector (in a particular direction). Geometric fragmentation arose 
in this model due to the stacking of 2D lattices containing \emph{confining 
magnetic excitations}, while the interlayer fluxes are frozen. The mobility 
of the magnetic excitations are highly restricted, as we show below, and these
are the fractons of this model. We begin by identifying the winding sectors in 
2D lattices containg fractons, and study how their presence influences the dynamics 
of the stacked lattices. 

 For the 2D QDM with periodic boundaries and of size $L \times (nL + 2)$, where 
$n \in \mathbb{N}_{0}$, we find that the winding sector characterized by 
$|W^{2D}_y| = L - 1$ and $|W^{2D}_x| = n$ exhibits a highly nontrivial 
structure and supports only severely constrained dynamics. By symmetry, an analogous 
result holds for sector with exchanged indices. The topology of the electric fluxes 
is such that configuration in this sector contains exactly two flippable plaquettes
sharing a common link (see for example Fig 4 (left)). 
Starting from either of the two flippable plaquettes, which are the two fractons
in this case, the kinetic term of the Hamiltonian acting on any initial state leads 
to a sequence of states where one of the flippable plaquettes in the pair 
move diagonally. Due to periodicity, this fracton pair, cycles through the entire 
lattice and returns to the initial state after exactly $L \times (nL + 2)$ flips.  
As a result, this 2D winding space class is especially interesting due to its atypical 
structure and the emergence of strongly frustrated dynamics. This structure naturally 
influences the dynamics of the 3D model.

\begin{figure}[h!]
  \centering
    \includegraphics[width=\linewidth]{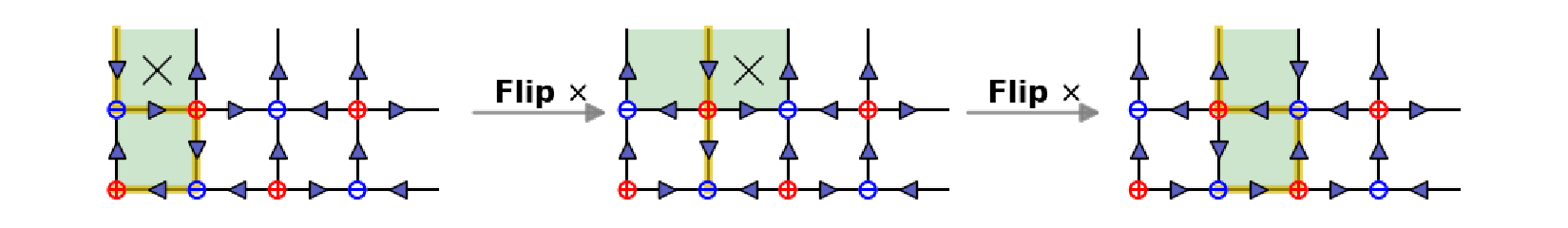}
    \caption{\label{fig:fractons_4x2} Inchworm motion of the flippable plaquettes 
    across the entire 2D plane, the two flippable plaquettes are fractions with
    limited mobility.}
\end{figure}

The construction of the states in the aforementioned winding spaces follows the 
specific structure described next and gives rise to frustrated dynamics. Understanding 
the construction helps to quantify the allowed dynamics analytically. Consider the
specific example of \( n = 0 \) and \( L = 4 \), which is the \( 4 \times 2 \) lattice, 
shown in the left panel of \cref{fig:fractons_4x2}. We begin with a configuration in which 
an S-shaped flux loop (in thick orange) is drawn at the lower left edge which connects
through periodic space by traversing the lattice in both the x- and y-directions. Two 
successive plaquette flips on plaquettes marked by a cross in left and middle of 
\cref{fig:fractons_4x2} return the system to the original configuration, translated 
spatially in x and y direction each by a single lattice spacing. As a result, the entire 
subspace is spanned by only two distinct states, up to a spatial translation of fluxes 
consistent with the background charge distribution.

\begin{figure}[h!]
  \centering
  \includegraphics[width=\linewidth]{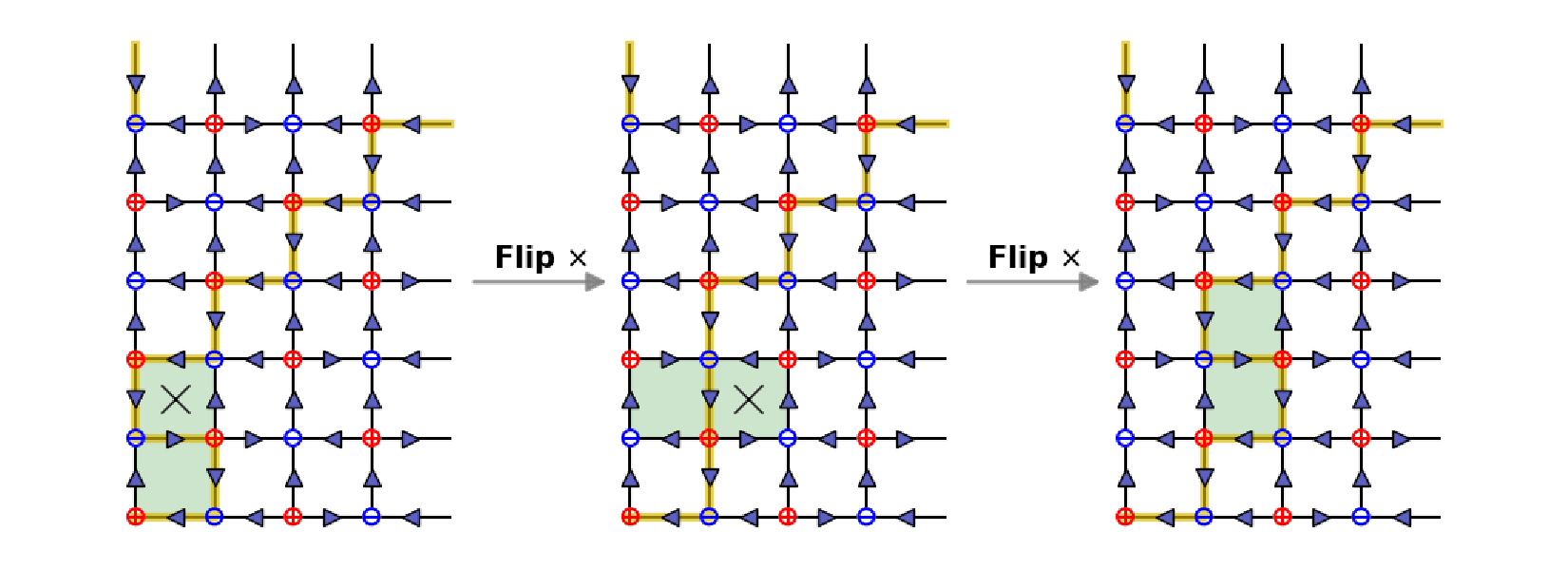}
  \caption{\label{fig:fractons_4x6} The inchworm motion of fractons on the $4\times6$ lattice.}
\end{figure}

It is instructive to examine now the case \( L = 4 \) and \( n = 1 \) in more detail. 
This choice corresponds to a \( 4 \times 6 \) lattice. As a representative case (among 
four symmetry-related winding spaces), we focus on the winding sector \( W^{2D}_{y} = 1 \), 
\( W^{2D}_{x} = -1 \). The leftmost panel in \cref{fig:fractons_4x6} is the starting 
state of our construction of this winding space. In orange, one can see a flux loop 
connected through periodic two-dimensional space that, except for an S-shape in the 
lower left corner, has a \emph{staircase} structure through the lattice. The fluxes 
forming the flux loop fully constrain the rest of the configuration. Given the periodic 
boundary conditions, the winding numbers, and the staggered charges, this is the only 
compatible flux pattern on the remaining links. Flipping the plaquette marked with a 
black cross results in the state in the middle panel of \cref{fig:fractons_4x6}. One 
can see that there remain 2 flippable plaquettes sharing a link, but now they are
horizontal neighbours. Flipping the plaquette marked with a black cross results in 
the state shown on the most right most panel \cref{fig:fractons_4x6} where flippable
plaquettes are again vertical neighbours, and simply related to the state in the leftmost
panel via a lattice translation in the $x$- and $y$-direction each. Subsequent flips lead
to a staircase-like movement of the fracton (the flippable plaquette pair) through the 
2D plane, which resembles a worm crawling along a line, and hence we dub it as the 
\emph{inchworm} motion. The whole winding subspace can thus be described by two states 
up to a translation. Constructing the states for $n>1$ involves addition of a \emph{staircase}
of directed flux to be able to close the flux loop. For $L=4$ and $n=2$, this is 
illustrated in \cref{sec:4x10_fracton_winding_sector}).

It is now clear how for arbitary $L$ and $n$ this construction can be extended:
we start with the S-shape flux line in the lower left corner and close the flux-loop 
by drawing $n$ staircase-like flux lines through the lattice. The remaining fluxes 
are fixed by this construction due to the Gauss's law and the 2D winding numbers (example for $n=2$ and $L=4$ in \cref{sec:4x10_fracton_winding_sector}). 
The three additional winding sectors, related by 
symmetry, can be easily constructed. These sectors are obtained by either starting 
with an inverted S-shape flux line at the lower left corner (corresponding to a start 
of the flux line at a positive charge), by initiating the S-shape one site shifted 
in the $x$-direction (start from the negative charge), or by combining both modifications 
(i.e., a shift and an inverted S-shape). The inverted configuration results in a 
staircase-like motion that is also inverted (mirrored) about the \( y \)-axis, 
while initiating the construction from the negative charge leads to a reversal 
in the direction of the flux line. Illustrative examples of these constructions 
for the case \( L = 4 \), \( n = 1 \) are provided in \cref{sec:sym_fractons}. 
We emphasize the severely restricted mobility of these excitations: even though
the parent system has a three-dimensional extent, these excitations can only propagate
along a single line cutting across the diagonal.

The winding sectors discussed above exhibit highly unconventional dynamics, characterized 
by the emergence of fractons. While the behavior of these excitations is already of 
considerable interest in the purely two-dimensional setting, we do not pursue a 
detailed analysis in this work. Instead, we take a step further by connecting these 
findings to our results on fragmentation in the three-dimensional case and the associated 
conservation of winding numbers. In the fragmented regime, the 3D lattice effectively 
consists of decoupled stacks of 2D layers, within which the 2D winding numbers remain 
conserved. Consequently, the system's evolution is confined within these 2D winding sectors. 
We identify two distinct classes of fragmented subspaces, each exhibiting a 
characteristic dynamics as the system is evolved in real-time: 

\begin{itemize}
\item The \textit{fractonic} fragmentation class is one where at least one fragment 
of the winding subspace can be understood in terms of the motion of planar fractons. 
Certain fragments in this category can be constructed by stacking 2D lattices whose 
basis states are related by the constrained motion of planar fractons, while there 
might exist other fragments which contain at least one 2D layer where the dynamics 
are not governed by planar fractons. The fragments that can be described by layers 
of moving fractons have $E_{\rm pot}(t)=\mathrm{const}$ and $E_{\rm kin}(t)=0$ for 
all starting states, because the total number of flippable plaquettes cannot change. 
Due to the constrained motion, we observe even stronger resistance to thermalization 
and large oscillations in entropy and fidelity in time. We consider cases where some 
2D-layers are frozen in a fragment and all the others are constructed by 2D planes 
hosting fractons. Since the frozen layers do not add any dynamics, the entire dynamics
are from the layers with fractions.
\item The \textit{non-fractonic} fragmentation class is defined by the absence
of fragments with \emph{only} fracton-dominated 2D layers. The dynamics in each fragment 
is constructed by 2D layers, such that at least one layer does not allow fractons. 
Thus, all ways to stack the 2D subspaces have at least one subspace without a constant 
number of flippable plaquettes. We want to highlight that this class can, but not necessarily 
has, to support fractons. Here, the $E_{\rm pot}(t)$ and the $E_{\rm kin}(t)$ are not 
constant in time for any of the starting states. This class also shows resistance to 
thermalization.
\end{itemize}

We now illustrate the behavior of each of the fragmentation classes 
concretely through examples. In each example, we show how the 3D winding 
numbers can be satisfied by stacking 2D lattices, and then show results
of real-time dynamics initialized from a product state in the respective
fragmented sector. Our numerical examples focus on lattices that extend 
over two sites in two directions and four sites in the remaining direction. 
For these system sizes, full exact diagonalization is feasible. In such 
lattices, the maximal winding number magnitude depends on the lattice 
extent: if a direction spans two sites, the maximal flux is 
$\max(|W^{3D}_{i}|) = 4$, whereas if it spans four sites, the maximum is 
$\max(|W^{3D}_{i}|) = 2$. We construct states according to the conventions 
described in \cref{sec:fragmentation_doped_lattice}, and we choose the 
maximal flux to be along the $z$-direction. Depending on whether the maximal 
extent of the lattice is along $z$, the construction involves stacking either 
four $2 \times 2$ lattices (corresponding to $|W^{3D}_{z}| = 2$), or two 
$4 \times 2$ lattices (corresponding to $|W^{3D}_{x}| = 4$). The available 
winding sectors for each lattice geometry are summarized in \cref{table:2x2} 
and \cref{table:4x2}, where we also indicate whether the corresponding 2D 
winding sector contains fractons.

Let's start with a very special example of a fractonic fragmentation class is the 
$(W_x,W_y,W_z)=(0,0,2)$ sector of the $2\times2\times4$ lattice. There are 
36 states that have no flippable plaquettes (each a frozen subspace), one 
big fragmented subspace that consists of 256 states, and 24 small fragmented 
subspaces each consisting of 16 states. This sector is constructed by stacking 
four $2\times2$ lattices in in the $z$-direction such that
\begin{equation}
W^{3D}_{x}=\sum_{i=1}^{4} W^{2D}_{x,i}=0 \text{ and } W^{3D}_{y}=\sum_{i=1}^{4} W^{2D}_{y,i}=0
\label{eq:3D_W_002}    
\end{equation}
There are multiple ways to do this:

\begin{itemize}
\item[1.] Four $(W^{2D}_{x},W^{2D}_{y})=(0,0)$ 2D winding sector lattices 
          stacked above each other. This corresponds to the big fragment with 256 states.
\item[2.] Two $(W^{2D}_{x},W^{2D}_{y})=(0,0)$ winding sector lattices stacked with one 
          $(W^{2D}_{x},W^{2D}_{y})=(0,1)$ winding sector lattice and one 
          $(W^{2D}_{x},W^{2D}_{y})=(0,-1)$ winding sector lattice. There are 12 ways to 
          stack these four planes above each other. Each sector has 16 states coming from
          the ones in the 2D planes with zero winding. 
\item[3.] Two $(W^{2D}_{x},W^{2D}_{y})=(0,0)$ winding sector lattices and one 
          $(W^{2D}_{x},W^{2D}_{y})=(1,0)$ winding sector lattice and 
          one $(W^{2D}_{x},W^{2D}_{y})=(-1,0)$ winding sector lattice. 
          There are 12 ways to stack these four planes above each other, each with 16 states.
\item[4.] Finally, there are 36 different ways to stack $(W^{2D}_{x},W^{2D}_{y})=(1,0)$ and 
        $(W^{2D}_{x},W^{2D}_{y})=(-1,0)$, $(W^{2D}_{x},W^{2D}_{y})=(0,-1)$ and 
        $(W^{2D}_{x},W^{2D}_{y})=(0,1)$ 2D planes above each other and fulfill 
        \cref{eq:3D_W_002}. These are not interesting because in these are only 2D 
        lattices without flippable plaquettes stacked above each other.
\end{itemize}

From the above classification, it follows that taking an initial state in any of these
sectors, and then doing an unitary time-evolution maintains a constant potential energy. 
Because the total energy must also be constant in time, their kinetic energies are constant 
in time as well. All non-frozen fragments are completly described by the dynamics of fractons.

\begin{figure}[h!]
  \centering
  \begin{minipage}[b]{0.24\textwidth}
    \includegraphics[scale=0.7]{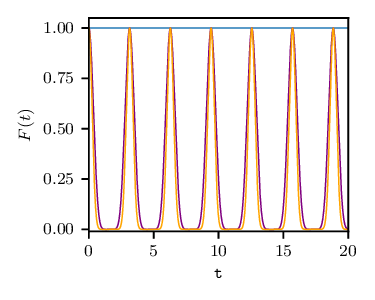}
  \end{minipage}
  \begin{minipage}[b]{0.24\textwidth}
   \includegraphics[scale=0.7]{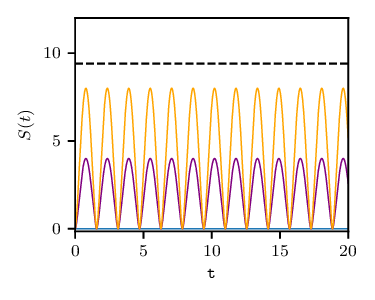}
  \end{minipage}
  \begin{minipage}[b]{0.24\textwidth}
    \includegraphics[scale=0.7]{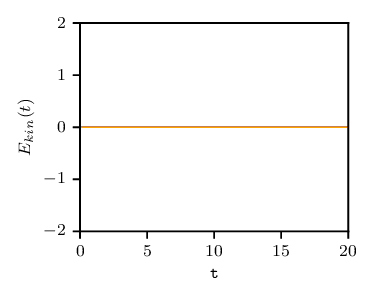}
  \end{minipage}
  \begin{minipage}[b]{0.24\textwidth}
   \includegraphics[scale=0.7]{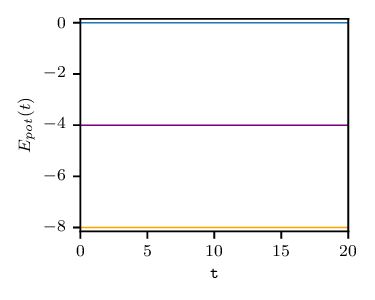}
  \end{minipage}
  \hfill
  \caption{Real-time dynamics of the \textbf{fractonic fragmentation class}, for the case of the
  $(0,0,2)$ sector of the $2\times2\times4$ lattice. The figures plot fidelity, Shannon entropy,
  the kintetic and the potential energy, from left to right.}
  \label{fig:subspace_2} 
\end{figure}

 \cref{fig:subspace_2} summarizes the dynamics for our four measures of 
interest. As expected, the states without flippable plaquettes corresponding 
to construction 4 (blue line) do not change in time, their fidelity is 
always equal to 1, and their entropy, kinetic energy, and potential energy 
are always equal to 0. The behaviour in time for all four measures of the 
states within the winding sector fragments is identical regardless of the 
starting state. The small subspaces (solid purple line) and the big subspace 
(solid orange line) show oscillations in the fidelity and the entropy each 
with a constant amplitude and period, so thermalization is evaded. All 
fragments are described by the dynamics of planar fractons and the fragments 
only differ in the amount of frozen 2D subsystems.

Another example of the fractonic fragmentation class is the 
$(W_x,W_y,W_z)=(1,0,4)$ sector of the $2\times4\times2$ lattice.
In this winding sector, there are two large fragmented spaces with 
128 states each, and two small ones of 8 states each. The sector 
is constructed by stacking two $2\times4$ lattices in the 
$z$-direction such that 
\begin{equation}
W^{3D}_{x}=\sum_{i=1}^{2} W^{2D}_{x,i}=1 \text{ and } W^{3D}_{y}=\sum_{i=1}^{2} W^{2D}_{y,i}=0 .
\label{eq:3D_W_410}    
\end{equation}
The fragmented subspaces are formed specifically by stacking:
\begin{itemize}
\item[1.] One $(W^{2D}_{x},W^{2D}_{y})=(1,0)$ plane, and one $(W^{2D}_{x},W^{2D}_{y})=(0,0)$ plane. 
There are two ways to do this stacking.
\item[2.] One $(W^{2D}_{x},W^{2D}_{y})=(2,0)$ and plane, and one $(W^{2D}_{x},W^{2D}_{y})=(-1,0)$ 
plane, and again there are two ways to do the stacking.
\end{itemize}

\begin{figure}[h!]
  \centering
  \begin{minipage}[b]{0.24\textwidth}
    \includegraphics[scale=0.7]{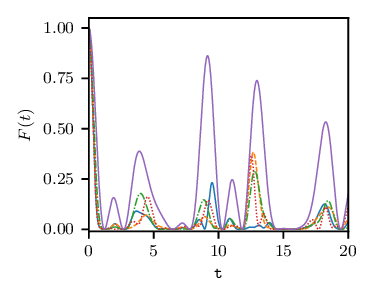}
  \end{minipage}
  \begin{minipage}[b]{0.24\textwidth}
   \includegraphics[scale=0.7]{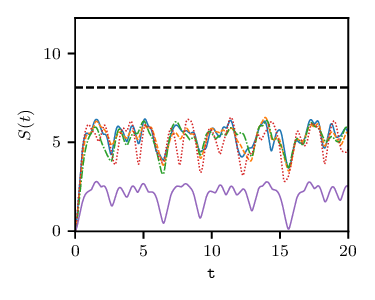}
  \end{minipage}
  \begin{minipage}[b]{0.24\textwidth}
    \includegraphics[scale=0.7]{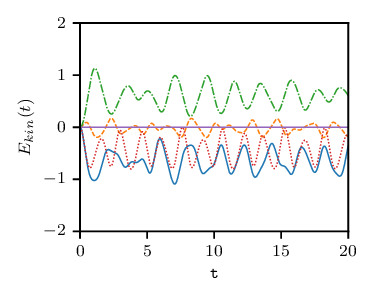}
  \end{minipage}
  \begin{minipage}[b]{0.24\textwidth}
   \includegraphics[scale=0.7]{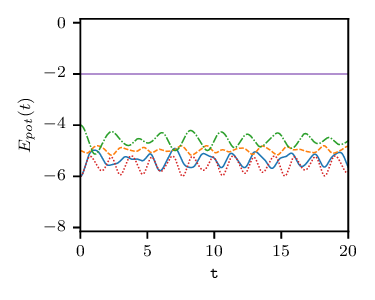}
  \end{minipage}
  \hfill
  \caption{Real-time dynamics of the \textbf{fractonic fragmentation class} (example 2). 
  Once again, from left to right, the behaviour of the fidelity, the Shannon entropy, kinetic 
  and potential energies are shown. The athermal nature of the evolution is evident in the 
  persistent oscillations that do not decay with time. 
  \label{fig:subspace_4} 
  }
\end{figure}

The results are shown in \cref{fig:subspace_4}. We observe four distinct types of behaviours 
in the large subspaces whereas in the small subspaces 
(purple line), the behaviour is independent of the starting state. In all subspaces we 
observe oscillations in the entropy and the fidelity: the entropy does not approach a steady 
state and the overlap with the initial wavefunction does not settle at zero. These oscillations 
are larger in the two small fragmented spaces. None of them thermalize according to ETH ---
note that for all of them the fidelity is significant even for very late times, and the Shannon
entropy is far from the one a maximally entangled state would possess. 
In the large subspaces, the expectation values of the kinetic and potential energy fluctuate 
in time, whereas in the small subspaces the expectation value of the kinetic energy is zero 
and the potential energy is constant in time. These are the fragments where all 2D subsystems 
are described by planar fractons and their dynamics.

\begin{figure}[h!]
  \centering
  \begin{minipage}[b]{0.24\textwidth}
    \includegraphics[scale=0.7]{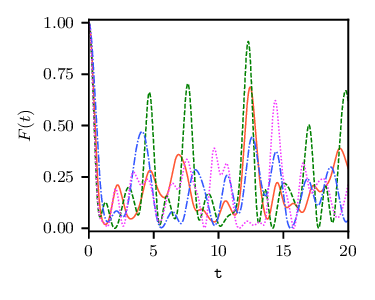}
  \end{minipage}
  \begin{minipage}[b]{0.24\textwidth}
   \includegraphics[scale=0.7]{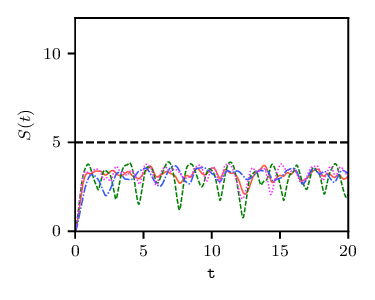}
  \end{minipage}
  \begin{minipage}[b]{0.24\textwidth}
    \includegraphics[scale=0.7]{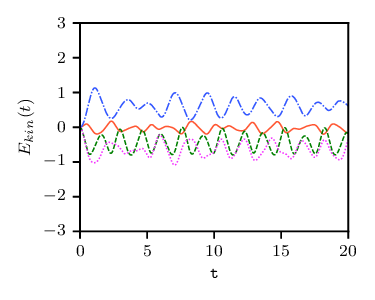}
  \end{minipage}
  \begin{minipage}[b]{0.24\textwidth}
   \includegraphics[scale=0.7]{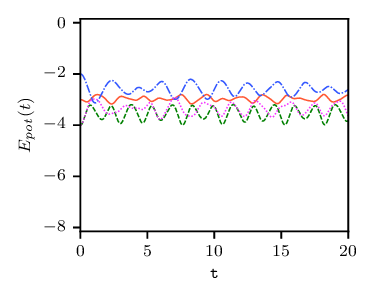}
  \end{minipage}
  \hfill
  \caption{Real-time dynamics for the \textbf{non-fractonic fragmentation class}.
  From left to right, the behaviour of the fidelity, the Shannon entropy, kinetic 
  and potential energies are shown. Even though fractons do not dominate this class
  of fragments, the persistent oscillations do not decay with time indicating athermal
  behaviour.
  \label{fig:subspace_3} 
  }
\end{figure}

An example winding space that is part of the non-fractonic fragmentation class is 
given by the $(W_x,W_y,W_z)=(0,1,4)$ sector. We want to highlight again that this is 
still a fragmented sector, but it does not have fragments completely dominated by 
fractons. This sector consists of two different fragmented subspaces each consisting 
of 16 states. This sector is constructed by stacking two $2\times4$ lattices in the 
$z$-direction such that
\begin{equation}
W^{3D}_{x}=\sum_{i=1}^{2} W^{2D}_{x,i}=0 \text{ and } W^{3D}_{y}=\sum_{i=1}^{2} W^{2D}_{y,i}=1
\label{eq:3D_W_401}    
\end{equation}
The only way to obtain these winding numbers is by stacking one plane with 
$(W^{2D}_{y}, W^{2D}_{z}) = (0,0)$ and one lattice plane with 
$(W^{2D}_{y}, W^{2D}_{z})=(0,1)$. There are obviously two different possibilities 
to do this construction.
We then observe four different evolution behaviors in each of the fragments 
depending on the starting state. Because the 2D subspaces that are stacked above each 
other do not have a constant number of flippable plaquettes, the potential energy is not 
constant in time, and thus the kinetic energy 
will change in time as well. No fragment is described by planar moving fractons stacked above 
each other. The four measures we computed are shown in \cref{fig:subspace_3}. All time-evolved states 
show fluctuations in the fidelity and the entropy. The entropy does not approach a steady value and 
the fidelity does not continuously go to zero. In this winding space, the expectation values of the 
kinetic and the potential energy are not constant, but fluctuate in time. The behaviour is not 
independent of the starting state. Still we do not observe full thermalization and see continued 
fluctuations in the entropy and fidelity in time, unlike what is predicted by ETH.

\subsubsection{Non-fragmented subspaces}
\begin{figure}[h!]
  \centering
  \begin{minipage}[b]{0.24\textwidth}
    \includegraphics[scale=0.7]{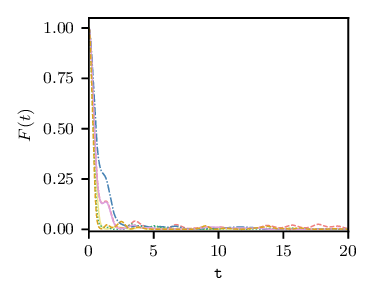}
  \end{minipage}
  \begin{minipage}[b]{0.24\textwidth}
   \includegraphics[scale=0.7]{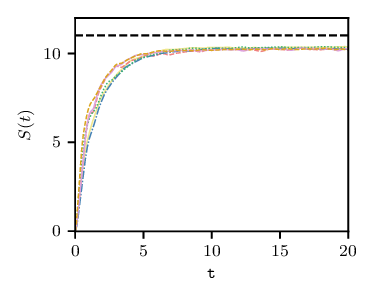}
  \end{minipage}
  \begin{minipage}[b]{0.24\textwidth}
    \includegraphics[scale=0.7]{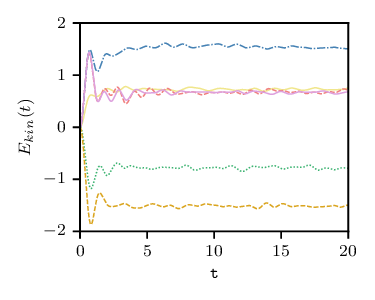}
  \end{minipage}
  \begin{minipage}[b]{0.24\textwidth}
   \includegraphics[scale=0.7]{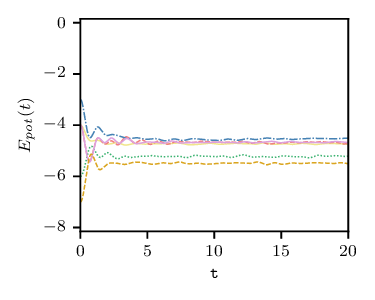}
  \end{minipage}
  \hfill
  \caption{Real-time dynamics of \textbf{non-fragmented subspaces}\label{fig:subspace_1}: 
  clear indications of saturation observed in overlap fidelity, Shannon entropy, kinetic 
  and potential energies (from left to right) as the system evolves unitarily in time.}
\end{figure}

We expect the non-fragmented subspaces to thermalize. As an example of a non-fragmented 
subspace of the $2\times2\times4$ lattice, we choose the $(W_x, W_y, W_z)=(2,2,1)$ subspace. 
This subspace does not have maximum flux in any direction and consists of 2084 states that are 
connected to each other by the Hamiltonian. We show the defined measures for nine random 
states of this subspace in \cref{fig:subspace_1}. The fidelity goes to nearly zero in very fast
and does not increase, and the entropy rises until it saturates near the maximal entropy value
(indicated as dotted black line), which indeed indicates that the system thermalizes. 

\subsection{Analytical solutions in the fractonic fragments}\label{sec:analytical_solution}
Here we give analytical solutions of the eigenstates and the eigenvectors 
for fragments that are completely dominated by planar fractons. 
We begin by investigating the 2D subspaces, because then 
\cref{eq:stacked_2D_lattice state} and \cref{eq:amiltonian_on_stacked_2D} will 
make it trivial to construct the 3D eigenstates and eigenvalues from those in 2D. 
Consider a lattice $L_x \times L_y=L \times (nL + 2)$, with 
$n \in \mathbb{N}_{0}$, in a winding sector defined by $|W^{2D}_y| = L - 1$ and $|W^{2D}_x| = n$. The subspace has $N=L_{x}\cdot L_{y}$ flux 
states, and each flux state has two flippable plaquettes with average potential energy 
$\alpha=-2\frac{\lambda}{J}$. Each flux state can is connected by the Hamiltonian to exactly two other flux states. 
By choosing the right order of the flux states, this results in the following matrix 
representation of the Hamiltonian:
\begin{equation}
\label{eq:Hamiltonian_matrix_fractons}
H^{2D}_{\rm fracton} = \alpha\ \mathbb{I}_{N\times N} + C_{N},
\end{equation}
where $\mathbb{I}_{N\times N}$ is the identity matrix and $C_{N}$ the adjacency matrix 
of the undirected cycle graph of N nodes
\begin{equation}
\label{eq:Hamiltonian_matrix_fractons}
C_{N} = \begin{bmatrix}
0 & 1 & 0 & 0 & \cdots & 0 & 1 \\
1 & 0 & 1 & 0 & \cdots & 0 & 0 \\
0 & 1 & 0 & 1 & \cdots & 0 & 0 \\
0 & 0 & 1 & 0 & \cdots & 0 & 0 \\
\vdots & \vdots & \vdots & \vdots & \ddots & \vdots & \vdots \\
0 & 0 & 0 & 0 & \cdots & 0 & 1 \\
1 & 0 & 0 & 0 & \cdots & 1 & 0
\end{bmatrix}.
\end{equation}
Because the identity matrix commutes with every matrix (and hence also with $C_{N}$),
it is sufficient to diagonalize $C_N$ to calculate the eigenenergies and eigenstates. 
The identity matrix introduces an absolute shift by $\alpha$ in the eigenenergies. 
It is a well known fact (e.g. \cite{cycle_graph_eigenspectrum}) that the adjacency 
matrix of the cycle graph is circulant with eigenvectors given by the discrete Fourier basis:
\begin{equation}
\label{eq:eigenstates_2D}
 \ket{\lambda_j}=\frac{1}{\sqrt{N}}(1, \omega^j,\dots,\omega^{(N-1)j})^T
\end{equation}
with $\omega = e^{2\pi i/N}$, $j=0,...,N-1$ and the eigenvalues are:
\begin{equation}
\label{eq:eigenvalues_2D}
\lambda_j = 2\cos(2\pi j / N).
\end{equation}
Therefore, the eigenenergies are $e_{j}=\alpha + \lambda_j$.

It is now trivial to construct the eigenspectrum and eigenvectors of the fracton-dominated fragments. Given $L_{z}$ number of stacked $L_{x}\times L_{y}$ 2D planes
dominated by fractons, the eigenvectors of the effective Hamiltonian 
\cref{eq:amiltonian_on_stacked_2D} are all possible combinations of the eigenstates 
of the 2D subsystems: 
\begin{equation}
\label{eq:eigenstates_3D}
 \ket{E_{\{j\}}}=\bigotimes_{l=1}^{L_z} \ket{\lambda_{j_{l}}}
\end{equation}
and the eigenenergies are the sum of the eigenenergies of the chosen eigenstate 
for each combination,
\begin{equation}
\label{eq:eigenenergies_3D}
 E_{\{j\}}=\sum_{l=1}^{L_z}e_{j_{l}}.
\end{equation}
The actual eigenvalues clearly depend on $N$. For stacking $2\times2$ planes, $N=4$,
which gives rise to the beating motion in \cref{fig:subspace_2}. Other examples
such as \cref{fig:subspace_4} have larger values of $N$ (e.g. $N=8$ for the purple
line in the figure) and give rise to more frequencies. The crucial point is that
no matter how large the lattices are, such frequencies are always present leading
to athermal behaviour.

\section{Conclusion and Outlook} \label{sec:outlook}
 While paradigms behind static properties of strongly correlated matter have acquired a
 maturity both in condensed matter and high-energy physics (even though novel phenomena
 are frequently reported), the analogous state of the field of non-equilibrium physics
 is still in its infancy. Understanding fragmentation, its causes, and its consequences,
 is indeed among the most challenging questions in the field.

 Gauge theories possess constrained Hilbert spaces due to local symmetries, which are 
 not enough to render the models integrable. Exotic behaviour is obtained when matter
 and gauge fields strongly interact and cannot be treated non-perturbatively, and very 
 few analytical results can be obtained. In this article, we are able to construct a 
 limiting case of a global symmetry in the cubic dimer model where many more emergent 
 subsystem symmetries are generated. These cause a geometric fragmentation of the model 
 which can be shown to be in the class of \emph{weakly} fragmented systems using analytic
 techniques. Further, the certain fragments have excitations which are highly constrained
 to move in two spatial dimensions \emph{smaller} than the original dimension of the
 system. The presence and absence of these fractonic excitations produce different
 dynamical behaviour, which we have also characterized using numerics on small lattices.
 We emphasize that we have given results showing this geometric fragmentation that are 
 valid on large lattices, and indeed in the thermodynamic limit. For fragmentation
 sectors which are dominated by fractons, we provide analytical solutions for 
 eigenvalues and eigenvectors.
 
 Several general consequences are obvious from our study. In particular, all the 
 statements about the fragmentation in \cref{sec:fragmentation_doped_lattice} were made
 for the cubic dimer model, which uses the Gauss sector where we have doped the system 
 with staggered $Q = \pm 1$. These results are also true for the Gauss charge zero sector,
 $Q=0$. In that sector, again, maximal flux restricts the action of the Hamiltonian to 
 2D planes orthogonal to the maximal flux, and the difference is that these 2D lattices 
 do not have charges in this case. The construction of the fragmented subspaces and the 
 weak fragmentation of the $W^{3D}_{x}=W^{3D}_{y}=0$ sector for all choices of $L_{x}$ 
 and $L_{y}$ for maximal flux in $L_{z}$ direction hold true for this charge sector, and
 can be generalized from our study. 

 It is easy to imagine various interesting directions of research emerging from our
 study. Are there other regimes possible in the pure gauge theory when other kinds of
 fragmentation emerge? It is also easy to imagine making the charges dynamical in a controlled
 fashion by endowing them with a large but finite mass. While winding numbers are not
 exact symmetries in the presence of charges, one can start with an initial state with
 a large flux and observe how the total flux evolves in time. This will clearly depend
 on the phase of the quenching Hamiltonian: a confined phase may try to keep the 
 fluxes around for longer, while in a Coulomb phase fluxes may dissipate by string
 breaking and recombination. In either case, exotic dynamical and athermal behaviour
 can be expected.

\section{Acknowledgments}
We would like to thank Arnab Sen for comments and discussions, particularly on the use
of entropy density as a diagnostic to distinguish between different fragmentation classes.
D.B. would like to thank STFC (UK) consolidated grant ST/X000583/1 and continued support 
from the Alexander von Humboldt Foundation (Germany) in the context of the research fellowship 
for experienced researchers. Research at Perimeter Institute is supported in part by the 
Government of Canada through the Department of Innovation, Science and Industry Canada and 
by the Province of Ontario through the Ministry of Colleges and Universities. This work 
was supported in part by the Helmholtz Association and the DLR via the 
Helmholtz Young Investigator Group ''DataMat''.

\bibliography{ref}

\appendix
\section{Supplementary Material}
\subsection{\label{sec:fragntation_proof}Fragmentation generalization proof}
\begin{figure}[h] 
	\centering
        \includegraphics{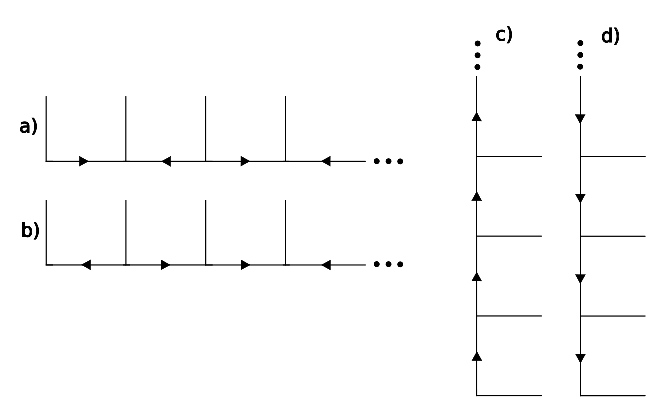}
	\caption{\label{fig:always existing} Orientation of the 
    \emph{flux lines} in planar 2D QDM states.} 
\end{figure}

\begin{figure}[h!]
  \centering
  \includegraphics[width=0.45\textwidth]{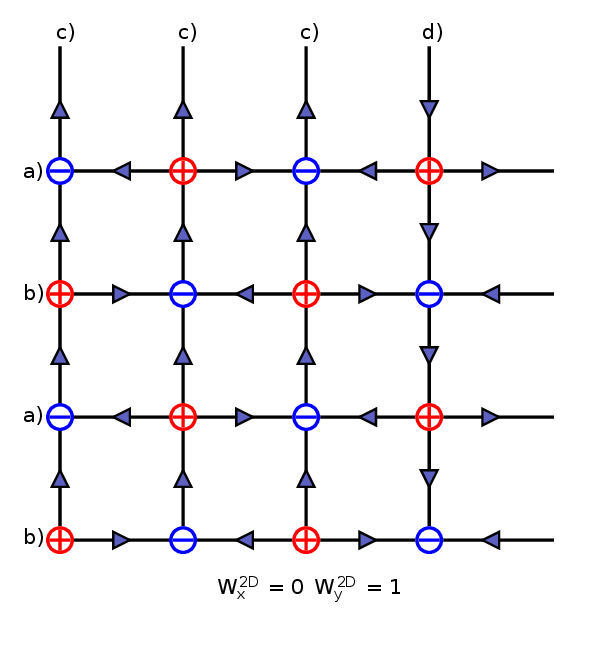}
  \hspace{1cm}
  \includegraphics[width=0.45\textwidth]{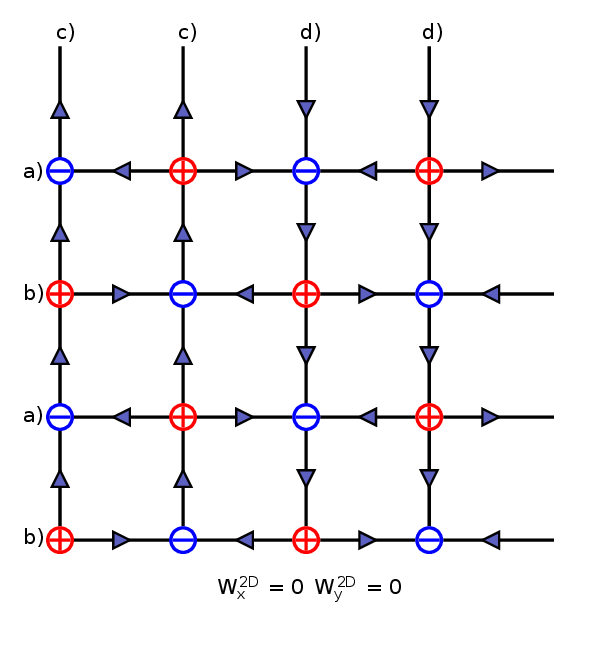}
  \caption{\label{fig:always existing2} Example of constructed 2D states
    which form the building blocks of the 3D system.} 
\end{figure}

In this subsection, we prove that for each lattice size there exist fragmented subspaces 
when the flux is maximal in one direction. Without loss of generality, for a lattice of size 
$L_{x} \times L_{y} \times L_{z}$, we assume that the maximum flux is positive, and  
$W^{\rm max}_{z}=(L_{x} \cdot L_{y})/2$. By the 
construction explained in the main text, all the states with 
$W^{\rm max}_{z}=(L_{x} \cdot L_{y})/2$ consist of $L_{z}$ planes of 2D dimer models 
of size $L_{x} \times L_{y}$ alternating between a positive and a negative charge as 
one moves in the positive $z$ direction. It is possible to construct states in each 
2D dimer lattice of any size $L_{x} \times L_{y}$ that contain non-trivial planar
dynamics. The \emph{flux lines} a) and b) presented in \cref{fig:always existing} 
extended to the length $L_{x}$ and any of the combinations of \emph{flux lines}
c) and d) extended to length $L_{y}$ can be combined to form a valid state. This 
produces a 2D dimer QLM state in winding sector $W^{2D}_{x}=0$ and 
$W^{2D}_{y}= n - m$, where $n$ is the number of c) flux lines, and 
$m$ is the number of d) flux lines. Which flux line a) or b) we start with 
determines whether it is a positive or a negative charge in the lower corner. 
This means we can construct one state in winding subspace in $W^{2D}_{x}=0$ and 
$W^{2D}_{y}\in\{-L_{x}/2, \cdots , 0 , \cdots ,L_{x}/2 \}$. Due to symmetry this means we can 
easily construct another state in $W^{2D}_{x}\in\{-L_{y}/2, \cdots, 0, \cdots, L_{y}/2 \}$ 
and $W^{2D}_{y}=0$. Examples of these constructions are shown in \cref{fig:always existing2} 
for a $4\times 4$ lattice.

To show that a winding subspace of a 3D lattice with maximal flux in one direction 
is fragmented, one has to show that \cref{3D_W} is fulfilled at 
least with two different combinations of planes. The above constructed states are 
useful to do this.

The winding space $(W^{3D}_{x},W^{3D}_{y})=(0,0)$ and maximal winding in z-direction
is always fragmented because we can construct two states with different combinations 
of 2D winding numbers out of the 2D Dimer planes (that exist for every lattice size)
and produce the correct 3D winding numbers, i.e. fulfill equations in \cref{3D_W}:

\begin{itemize}
    \item Construction 1:
        \begin{center}
        \begin{math}
            L_{z} \times \mathrm{xy{\text -}planes\; with} \ W^{2D}_{x}=0,\ W^{2D}_{y}=0
        \end{math} 
        \end{center}
        
    \item Construction 2:
        \begin{center}
        \begin{math}
           \frac{L_{z}}{2} \times \mathrm{xy{\text -}planes\; with} \ W^{2D}_{x}=0, \ W^{2D}_{y} = -1
        \end{math}  
        
        \begin{math}
            \frac{L_{z}}{2} \times \mathrm{xy{\text -}planes\; with} \ W^{2D}_{x}=0, \ W^{2D}_{y} = +1
        \end{math}
        \end{center}
\end{itemize}
The above construction guarantees the existance of at least a single state with the winding numbers listed in constructions 1 and 2.
This means that there are two different states that are not connected by the Hamiltonian, i.e. the subspace is fragmented.\\

The winding sectors satisfying the equations  
$W^{3D}_{x} = a \times d \quad \text{and} \quad W^{3D}_{y} = b \times e,$
where \( d \in \{-L_{y}/2, \ldots, 0, \ldots, L_{y}/2\} \) and \( e \in \{-L_{x}/2, \ldots, 0, \ldots, L_{x}/2\} \),  
are also fragmented. This fragmentation arises because multiple distinct combinations of 2D winding numbers can yield the same total 3D winding numbers. In this construction, the values of \( d \) and \( e \) determine the winding contributions of the individual 2D planes to the total 3D flux. The remaining \( z \)-slices—those not contributing directly to the winding—can be filled with trivial planes carrying winding numbers \( (W^{2D}_x, W^{2D}_y) = (0,0) \). The number of such planes is given by \( c = L_z - a - b \). Clearly, \( a \), \( b \), and \( c \) are nonzero integers, and at least one of \( d \) or \( e \) must be nonzero (i.e., \( d \ne 0 \), \( e \ne 0 \), or both).

\begin{itemize}
    \item Construction 1 follows the order where we first stack the planes with non-trivial 2D 
    winding in x-direction, then the ones with trivial winding, and then the ones with 
    non-trivial winding in y-direction:
    \begin{center}
    \begin{math}
    a \times \mathrm{xy{\text -}planes\; with} \ W^{2D}_{x}=d,\ W^{2D}_{y}=0 
    \end{math}
    
    then
    
    \begin{math}
    c \times  \mathrm{xy{\text -}planes\; with} \ W^{2D}_{x}=0,\ W^{2D}_{y}=0
    \end{math}
    
    then
    
    \begin{math}
    b \times \mathrm{xy{\text -}planes\; with} \ W^{2D}_{x}=0,\ W^{2D}_{y}=e
    \end{math}
    \end{center}

    \item Construction 2 simply flips the order followed in Construction 1
    \begin{center}
    \begin{math}
    b \times \mathrm{xy{\text -}planes\; with} \ W^{2D}_{x}=0,\ W^{2D}_{y}=e
    \end{math}    
    
    then
    
    \begin{math}
    c \times \mathrm{xy{\text -}planes\; with} W^{2D}_{x}=0,\ W^{2D}_{y}=0
    \end{math}
    
    then
    
    \begin{math}
    a \times \mathrm{xy{\text -}planes\; with} \ W^{2D}_{x}=d,\ W^{2D}_{y}=0 
    \end{math}
    \end{center}
\end{itemize}
These constructions show fragmentation because the order of stacking the planes is 
important if they are of different 2D winding sectors.

\subsection{ Fragmentation of other winding sectors}
 \label{sec:non_zero_winding_strong_fragmentation}
 Here we prove the following: \emph{The winding sector $(l,m,W^{\rm max}_{z})$ 
of the $L\times L \times L$ lattices is weakly fragmented for $L\to\infty$ }. 
 
 We follow the same approach as in the main text for $(0,0,W^{\rm max}_{z})$. Based on the 
construction of stacking \( L_z \) two-dimensional lattices along the \( z \)-direction, 
it follows directly that the largest fragment within the three-dimensional winding sector 
\( W^{3D}_{x} = l \mathrm{ \ and \ } W^{3D}_{y} = m \), under conditions of maximal flux 
in the \( z \)-direction, corresponds to stacking \( l \) copies of the \( W^{2D}_{x} =1\) 
and \(W^{2D}_{y} = 0 \), \( m \) copies of the \( W^{2D}_{x} =0\) and \(W^{2D}_{y} = 1 \)  
and \(L_{z}-l-m\) copies of the \( W^{2D}_{x} = 0 \mathrm{ \ and \ } W^{2D}_{y} = 0 \) 
sector on top of one another to form a total of $L_z$ layers.
This follows from the fact that the zero winding sector 
is the largest sector for all 2D lattices. As we have argued before, in the 
thermodynamic limit ( $L_{x}\to\infty$ and $L_{y}\to\infty$), one has $n_{0}\approx n_{1}$,
where $n_0$ is the number of states in the sector $(W^{2D}_{x},W^{2D}_{y})=(0,0)$ sector
while $n_{1}$ is the number of states in the sector where either $|W^{2D}_x|$ or $|W^{2D}_y|$
is 1 and the other is 0. It follows that $N_{\rm largest} \approx n_{0}^{L_{z}}$. 
The lower bound is the same as before  because our construction of $N_{\rm max}$ 
and therefore the lower bound (\cref{eq:upper}) is still valid.

For the construction of $N_{\rm min}$ we now have to distinguish two different cases: 
where \(l+m\) is even, and where \(l+m\) is odd. Let us start with \(l+m\) even. As before, 
our construction pairs the layers such that the winding numbers of $(L-l-m)/2$ pairs cancel 
out and $l$ planes add the flux in x-direction, and $m$ planes add the flux in y-direction. 
We first choose $l+m$ planes that add up to the total flux in the x and y directions, 
while the remaining planes are paired as described before. For each of the $(L-l-m)/2$ pairs, 
we have five free choices for one plane in the pair, while the other plane in the pair is 
fixed by the first choice. This construction results in $5^{(L-m-l)/2}$ ways to stack the 
planes. Using the approximation \( n_0 \approx n_1 \), the total number of states in the 
constructed subset scales as \( N_{\text{min}} \approx n_0^{L_z} \cdot 5^{(L_z-m-l)/2} \). 
This provides an explicit upper bound on $\mathcal{G}_{\rm lower}$, and by extension to 
$\mathcal{F}$, since the proposed choices are a subset of the possible ways to construct the 
states. Quantitatively, we have

\begin{equation}
    \label{eq:strong_frag_scaling_3}
  \mathcal{F} = \lim_{L\to\infty} \frac{N_{\rm largest}}{N_{\rm total}} 
  < \mathcal{G}_{\rm upper} =   \lim_{L\to\infty} e^{-(L-m-l)\frac{\ln(5)}{2}}
\end{equation}
The proof for \(l+m\) odd is the same except that to be able to pair the planes 
that do not add flux in x- and y-directions, we have to include one fixed plane 
that is in the zero flux sector. In this case we have $(L-l-m-1)/2$ plane-pairs,
and for each we have three free choices. The rest of the proof is equivalent and results 
in the scaling:
\begin{equation}
    \label{eq:strong_frag_scaling_3}
  \mathcal{F} = \lim_{L\to\infty} \frac{N_{\rm largest}}{N_{\rm total}} 
    < \mathcal{G}_{\rm upper} = \lim_{L\to\infty} e^{-(L-m-l-1)\frac{\ln(5)}{2}}
\end{equation}

This shows that the $W^{3D}_{x}=l$ and $W^{3D}_{y}=m$ sector with maximal flux in 
z-direction is weakly fragmented in the thermodynamic limit. As before do we note that in the
proof we have considered cases where we use 2D dimer planes for which $W^{2D}_{x,y}$ 
is 0, or 1 in the construction of the upper bound. We have verified that including more 
2D winding sectors and more complex combinations than plane pairs still leads to a 
weak fragmentation, but the correction term to the entropy density will disappear 
more slowly in the thermodynamic limit (still ${\cal O} (1/L^2)$).

\subsection{\label{sec:4x10_fracton_winding_sector}
Fractons in larger winding sectors ($n>1$): example $4\times10$ lattice}
\begin{figure}[h!]
  \centering
    \includegraphics[width=\linewidth]{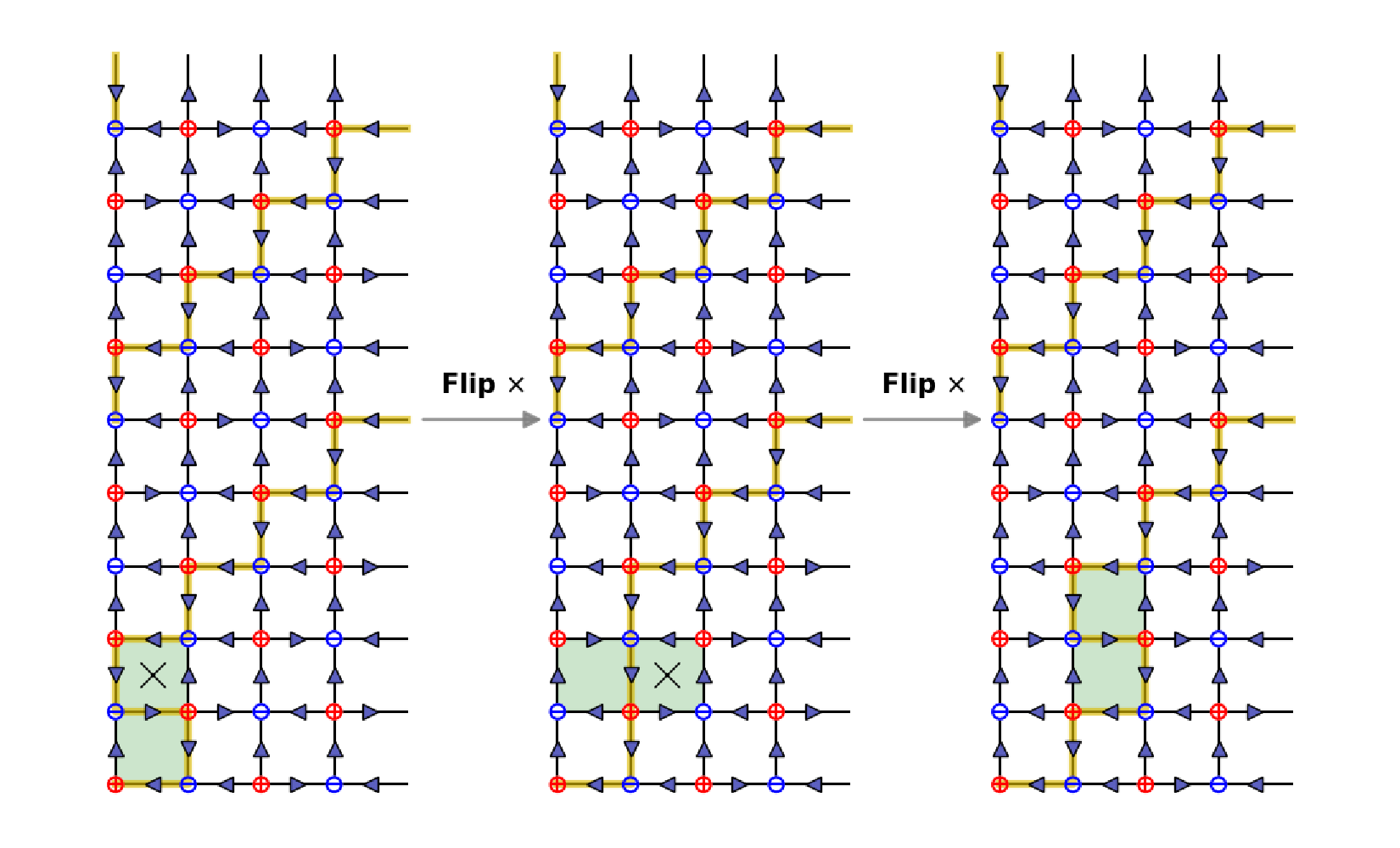}
  \caption{\label{fig:fractons_4x10} }
\end{figure}

It is also instructive to consider the case \( L = 4 \) and \( n = 2 \), 
focusing on the winding sector \( W^{2D}_{y} = 1 \), \( W^{2D}_{x} = -2 \),
as shown in \cref{fig:fractons_4x10}.

\subsection{\label{sec:sym_fractons}Symmetric fracton winding sectors}
\begin{figure}[h!]
  \centering
  \begin{minipage}[b]{0.3\textwidth}
    \includegraphics[width=\linewidth]{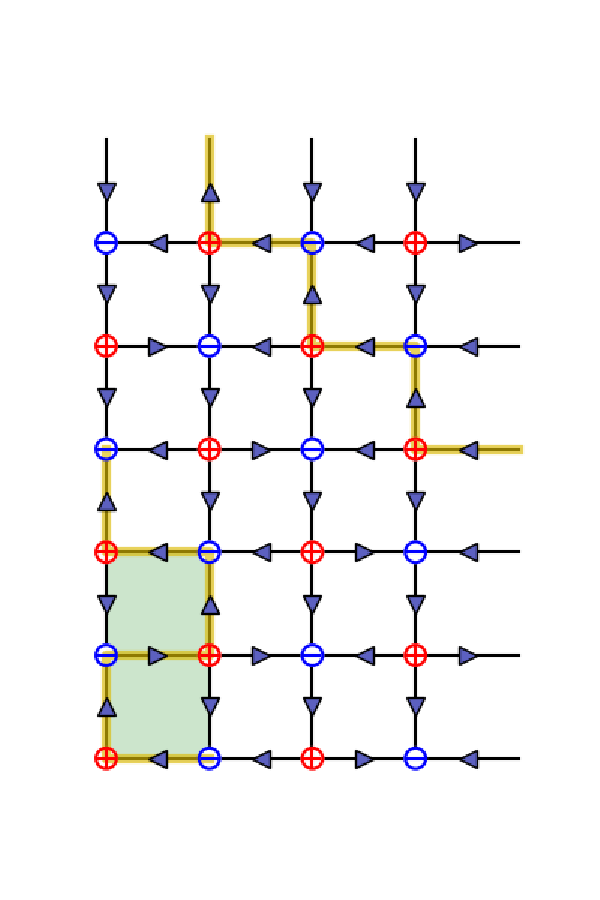}
    \subcaption{Construction start state $W^{2D}_{x}=-1$ and $W^{2D}_{y}=-1 $}
  \end{minipage}
  \begin{minipage}[b]{0.3\textwidth}
   \includegraphics[width=\linewidth]{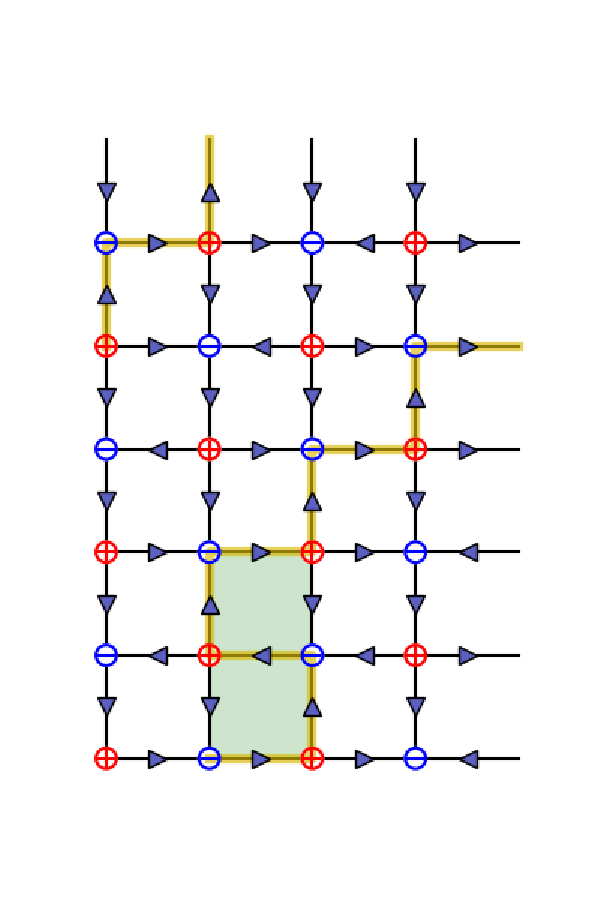}
   \subcaption{Construction start state $W^{2D}_{x}=1$ and $W^{2D}_{y}=-1 $}
  \end{minipage}
  \begin{minipage}[b]{0.3\textwidth}
    \includegraphics[width=\linewidth]{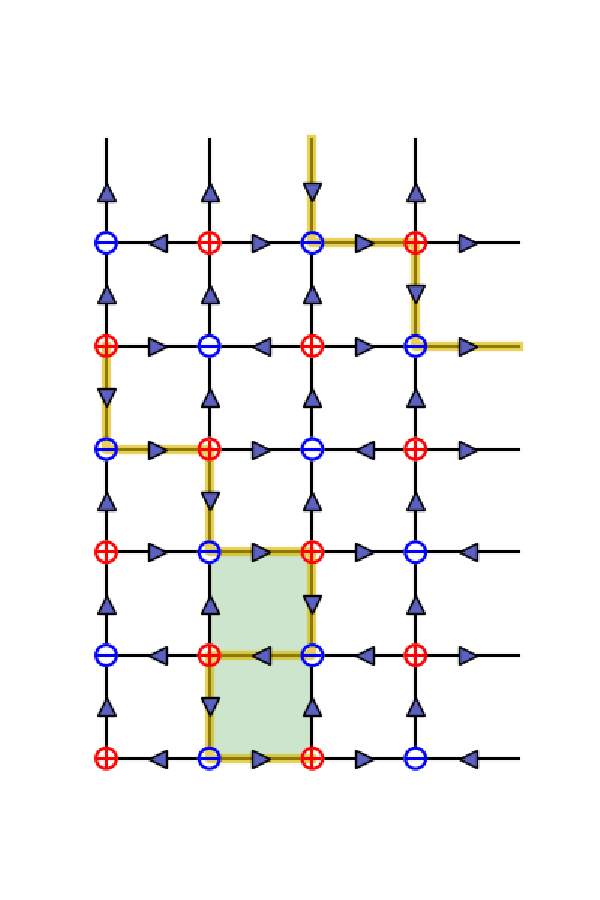}
    \subcaption{Construction start state $W^{2D}_{x}=1$ and $W^{2D}_{y}=1 $}
  \end{minipage}
\caption{}\label{fig:symmWind}
\end{figure}

 The construction of the symmetry related winding number sectors to the one
 discussed in the main text $(W_x^{2D},W_y^{2D})=(-1,1)$ is shown in 
 \cref{fig:symmWind}.
 
\end{document}